\documentclass[12pt,english]{article}

\usepackage{mathptmx}
\usepackage{mathrsfs}
\usepackage{geometry}
\usepackage{xcolor}
\usepackage{array}
\usepackage{bbding}
\usepackage{multirow}
\usepackage[fleqn]{amsmath}
\usepackage{amssymb}
\usepackage{amsthm}
\usepackage{graphicx}
\usepackage{natbib}
\usepackage{lscape}
\usepackage[hyphens]{url}
\usepackage[hidelinks]{hyperref}
\usepackage{comment}
\usepackage{bm}
\usepackage[title]{appendix}
\usepackage{longtable}
\usepackage{mathtools}
\usepackage{chngcntr}
\usepackage{authblk}
\usepackage{babel}
\usepackage{dsfont}
\usepackage{subcaption}
\usepackage{fancyhdr}
\usepackage{cleveref}
\usepackage{abstract}
\usepackage{algorithm}
\usepackage{algpseudocode}

\makeatletter
\allowdisplaybreaks
\date{}

\fancypagestyle{plain}{%
	\fancyhf{} % sets both header and footer to nothing
	
	\chead{\footnotesize \textcolor{gray}{\textit{\shortauthors -- \runningtitle}}}
	\rhead{\footnotesize \textcolor{gray}{\thepage}}
}
\makeatletter
\def\blfootnote{\xdef\@thefnmark{}\@footnotetext}
\makeatother

\makeatletter
\def\titlepageext{
	\begin{center}	
	{\parindent0pt
		\rule{0.9\textwidth}{1pt}
		\begin{minipage}[t]{0.25\textwidth}
			\small {\it Keywords:}\\
                \keywords
		\end{minipage}%
		\hspace{3mm}
		\begin{minipage}[t]{0.6\textwidth}
			\small \abstract
		\end{minipage}%
		
		\rule{0.9\textwidth}{2pt}
	}% restore indentation
	\end{center}

	\blfootnote{* Corresponding author. E-mail address: \href{mailto:\corresemail}{\corresemail}.}
}
\makeatother

\usepackage[mathlines, pagewise]{lineno}
\usepackage{etoolbox} %% <- for \pretocmd, \apptocmd and \patchcmd

\newcommand*\linenomathpatchAMS[1]{%
	\expandafter\pretocmd\csname #1\endcsname {\linenomathAMS}{}{}%
	\expandafter\pretocmd\csname #1*\endcsname{\linenomathAMS}{}{}%
	\expandafter\apptocmd\csname end#1\endcsname {\endlinenomath}{}{}%
	\expandafter\apptocmd\csname end#1*\endcsname{\endlinenomath}{}{}%
}

\expandafter\ifx\linenomath\linenomathWithnumbers
\let\linenomathAMS\linenomathWithnumbers
\patchcmd\linenomathAMS{\advance\postdisplaypenalty\linenopenalty}{}{}{}
\else
\let\linenomathAMS\linenomathNonumbers
\fi

\linenomathpatchAMS{gather}
\linenomathpatchAMS{multline}
\linenomathpatchAMS{align}
\linenomathpatchAMS{alignat}
\linenomathpatchAMS{flalign}

\nolinenumbers

\newtheorem{proposition}{Proposition}

\newtheorem{definition}{Definition}

\title{On Path-based Marginal Cost of Heterogeneous Traffic Flow for General Networks}

\def\shortauthors{Liu \& Qian}
\def\runningtitle{PMC of heterogeneous traffic flow}

\author[a]{Jiachao Liu}
\author[a,b,$\ast$]{Sean Qian}
\affil[a]{Department of Civil and Environmental Engineering, Carnegie Mellon University, Pittsburgh, PA, U.S.A}
\affil[b]{Heinz College, Carnegie Mellon University, Pittsburgh, PA, U.S.A.}

\def\corresemail{seanqian@cmu.edu}

\def\abstract{Path marginal cost (PMC) is a crucial component in solving path-based system-optimal dynamic traffic assignment (SO-DTA), dynamic origin-destination demand estimation (DODE), and network resilience analysis. However, accurately evaluating PMC in heterogeneous traffic conditions poses significant challenges. 
Previous studies often focus on homogeneous traffic flow of single vehicle class and do not well address the interactive effect of heterogeneous traffic flows and the resultant computational issues.
{This study proposes a novel but simple method for approximately evaluating PMC in complex heterogeneous traffic condition. The method decomposes PMC into intra-class and inter-class terms and uses conversion factor derived from heterogeneous link dynamics to explicitly model the intricate relationships between vehicle classes. Additionally, the method considers the non-differentiable issue that arises when mixed traffic flow approaches system optimum conditions. The proposed method is tested on a small corridor network with synthetic demand and a large-scale network with calibrated demand from real-world data. Results demonstrated that our method exhibits superior performance in solving bi-class SO-DTA problems, yielding lower total travel cost and capturing the multi-class flow competition at the system optimum state.}}

\def\keywords{Dynamic traffic assignment\\System Optimum\\Path Marginal Cost\\Heterogeneous Traffic Flow\\Subgradient}

\begin{document}
\maketitle
\titlepageext

\section{Introduction}
For general networks, Path-based Marginal Cost (PMC) measures the change in generalized path travel cost with respect to incremental change in class-specific path flow or link flow. It is the key to many network applications. For instance, it measures the stability and reliability of travel time (a particular realization of generalized cost) among network flows. It also helps derive gradients with respect to path/link flow in networks which are essential to estimate path/link flow (e.g. in dynamic origin-destination estimation problems). Path-based Marginal Cost is notoriously difficult to solve because the mapping from network flow to system cost is non-smooth and even non-continuous depending on the queuing models and the choice of flow or agent-based models in the network context. 

Another key application of PMC is to solve for system optimum. The system-optimal dynamic traffic assignment (SO-DTA) solves the time-varying traffic flows across a network that minimize the total generalized system cost, given the origin-destination (O-D) travel demand. The solution of SO-DTA usually serves as the benchmark for traffic system management and has been widely studied for decades. One possible approach to formulate SO-DTA is link-based in which base variables are time-dependent link flows \citep{merchant1978model,merchant1978optimality,carey1987optimal}. However, the link-based SO-DTA suffer from a few issues such as non-convexity \citep{CAREY1992127,merchant1978model,merchant1978optimality}, vehicle holding \citep{ziliaskopoulos2000linear,han2011complementarity,zhu2013cell} and the potential violation of First-in-first-out (FIFO) \citep{CAREY1992127}. Some recent study address these challenges, but link-based formulations are usually mathematically intractable for large network applications due to both high non-convexity and high dimension \citep{long2018dynamic,long2019link}. Another approach to formulate SO-DTA is path flow-based \citep{ghali1995model,peeta1995system,lo1999dynamic,ukkusuri2012dynamic}. The path-based formulation has linear constraints but the objective function of total travel cost is usually highly non-linear due to the fact that the travel costs are computed based on simulations of network flow dynamics. {The path-based formulation can be cast into a variational inequality problem (VIP) and solution algorithms for VIP formulation requires PMC evaluations \citep{shen2007path,qian2012system}, and the final system optimum (SO) state should ensure equalized PMCs across all paths and departure times.}

The PMC can be analytically evaluated through a dynamic network loading (DNL) process and no closed form is available due to the model complexity of DNL. Existing methods usually evaluate PMC by tracing a positive flow perturbation in the DNL based on the implicit assumption that the total travel cost is differentiable at all times with respect to path flow \citep{shen2007path,qian2011computing,qian2012system}. However,  it has been proved by \cite{zhang2020path} that the PMC is non-differentiable especially at optimum and sub-gradients should be used instead. {Another issue comes from the multi-class modeling in DNL where there coexist multiple vehicle classes (i.e., cars, trucks, buses, etc.) representing various traffic modes in transportation systems.} These vehicle classes have different {traffic flow characteristics} and influence each other in multi-class DNL \citep{qian2017modeling}, {and heterogeneity assumptions sometimes affect the existence and uniqueness of equilibrium solutions in bottleneck models \citep{ramadurai2010linear}.}
To the best of our knowledge, {none of existing research address this heterogeneous traffic interaction when calculating PMC for solving multi-class path-based SO-DTA.} {\cite{ma2020estimating} and \cite{liu2024modeling}} calculated PMC of cars and trucks based on separate cumulative curves of {two vehicle classes}, and no interaction is considered. Meanwhile, if considering the multi-class interaction, the conditions used to check if PMC is non-differentiable need to be compatible with the {heterogeneous} traffic dynamics.

To address these challenges, this study proposes {an} analytical approach to approximate PMCs in a multi-class DNL process, simultaneously considering the interaction between different vehicle classes and the non-differentiable issue. The main contributions can be summarized as follows: (1) This study proposes a novel approach for evaluating path marginal cost (PMC) within the context of heterogeneous, multi-class dynamic network loading. The proposed method captures the complex interplay between multiple vehicle classes {by modeling inter-class terms after PMC decomposition}. (2) This study resolves the non-differentiability issue inherent in multi-class PMC evaluation and establishes a theoretical connection with heterogeneous traffic flow theory. (3) The proposed PMC evaluation method facilitates enhanced performance in {solving multi-class SO-DTA}. Additionally, the method has potential for {calibrating network flow and travel time using probe speed data and network resilience analysis}.

The remaining paper is organized as follows. 
Section~\ref{sec:pre} presents the formulation of SO problem and revisit the multi-class DNL model in \cite{qian2017modeling} which is used throughout this study.
{Section~\ref{sec:model} presents the method of approximating PMC for bi-class traffic by decomposing PMC into intra-class and inter-class terms. We further present the standard Method of Successive Averages (MSA) algorithm to solve SO-DTA based on PMCs.}
Section~\ref{sec:experiment} compares and discusses experiment results on two networks of different sizes. 
Section~\ref{sec:conclusion} summarizes main findings of this study and gives future research directions.

\section{Preliminaries}
\label{sec:pre}
\subsection{Explaining PMC in Multi-class System Optimum}
Consider a general transportation network, denoted by $G = (N,A)$, which consists of a set of link $A$ and a set of node $N$. {In the network, there are $|I|$ categories of travelers represented by different vehicle classes $i\in I$.} $R\subset N$ and $S\subset N$ are sets of origin and destination nodes. {Between an OD pair $rs$ where $r\in R$ and $s\in S$}, there exists a route set for class $i$, denoted as $P^{rs}_i$. Let $T_a = [0,T]$ be the assign period and the network is initialized to be empty at time $t = 0$. $f^{rs}_{p,i}(t)$ denotes the path flow of class $i\in I$ departing at time $t\in T_a$ from the origin node $r\in R$, choosing path $p\in P^{rs}_i$ and heading to destination $s\in S$. The path flow $f^{rs}_{p,i}(t)$ is a function of departing time $t$ and each path flow has a corresponding travel cost, denoted by $c^{rs}_{p,i}(t)$ which is also a function of time $t$. {In the following equations, $f^{rs}_{p,i}(t)$ and $f^{rs}_{p,t,i}$ are used interchangeably and both denote time-dependent path flow. The same rule applies to other variables with time index.}
Generally, the travel cost consists of two parts, travel cost regarding actual travel time (time of value $\alpha$ times path traverse time {$w^{rs}_{p,i}(t)$}) and schedule delay penalty, {as shown in Equation~\ref{eq:ttc}}. The schedule delay penalty is a function {$\text{sd}(\cdot)$} depending on the difference between actual arrival time $t + w^{rs}_{p,i}(t)$ and desired arrival time {$t^{*}_{rs,i}$}. We use $\Delta t$ to denote the time difference $\Delta t = t + w^{rs}_{p,i}(t) - t^{*}_{rs,i}$. Travel time {$w^{rs}_{p,i}(t)$} can be obtained from any delay operator depending on the DNL process used.
\begin{equation}
\label{eq:ttc}
    c^{rs}_{p,i}(t) = \alpha w_{p,i}^{rs}(t) + f\left(\Delta t\right)
\end{equation}
The schedule delay penalty function is chosen to be piece-wise linear, depending on the $\Delta t$ and a pre-defined {punctuality threshold $\delta$ as follows}
\begin{equation}\label{eq:schedule}
    \begin{aligned}
        \text{sd}\left(\Delta t\right) = \begin{cases}
            \beta\left(-\Delta t - \delta\right) & \Delta t \in (-\infty, \delta)\\
            0 & \Delta t\in [-\delta, \delta]\\
            \gamma\left(\Delta t - \delta\right) & \Delta t\in (\delta, +\infty)\\
        \end{cases}
    \end{aligned}
\end{equation}
where $\beta$ and $\gamma$ are penalty coefficients for arriving early and late respectively. {The function {$\text{sd}(\cdot)$} can also depend on OD pair $rs$, path $p$ and vehicle class $i$, which can be written as $\text{sd}^{rs}_{p,i}(\cdot)$. Similar rules also apply to penalty coefficients and the punctuality threshold.}

The SO-DTA considering both departure time and route choice simultaneously can be formulated as
\begin{equation}
\label{continunous_so}
\begin{aligned}
    \min_{\left\{f^{rs}_{p,i}(t)\right\}_{r,s,p,t,i}}\textbf{TC}\left(\left\{f^{rs}_{p,i}(t)\right\}_{r,s,p,t,i}\right) &= \sum_{rs\in RS} \sum_{i\in I}\sum_{p\in P^{rs}_i}\int_{0}^T f^{rs}_{p,i}(t)c^{rs}_{p,i}(t) dt\\
    \text{s.t.}\quad \sum_{p\in P^{rs}_i}\int_{0}^{T}f^{rs}_{p,i}(t) dt &= q^{rs}_i,\quad \forall r,s,i\\
    f^{rs}_{p,i}(t) &\geq 0,\quad \forall r,s,p,t\\
\end{aligned}
\end{equation}
where the objective function is minimizing the total travel cost of all travelers in the system and the two constraints are flow conservation and non-negative conditions respectively. 

{Here we introduce the concept of Path Marginal Cost and related Link Marginal Cost, which will be used to reformulate the SO-DTA problem. Those concepts were previous discussed by \citet{shen2007path,qian2011computing,qian2012system,zhang2020path}}. 

\begin{definition}
    \textbf{Path Marginal Cost (PMC)} Path marginal cost, denoted by $\text{PMC}^{rs}_{p,t,i}$, is defined as the change in total system cost caused by {one unit flow perturbation} of class $i$ departing at time $t$ on path $p$ between OD pair $rs$.
\end{definition}
\begin{definition}
    \textbf{Link Marginal Cost (LMC)} The link marginal cost, {denoted by $\text{LMC}^{rs}_{a,t,i}$} is defined as the change in travel cost of link $a$ caused by {one unit flow perturbation} of class $i$ departing at time $t$ on path $p$ between OD pair $rs$.
\end{definition}
{According to \cite{shen2007path}, the PMC and LMCs have the following relationship.
\begin{equation}\label{eq:pmc_lmc}
    \text{PMC}^{rs}_{p,t,i}(\textbf{f}) = \text{sd}^{rs}_{p,i}(t) + \sum_{a\in A}\sum_{k}\frac{\partial (u_{a,i}(k)\omega_{a,i}(k,\textbf{f}))}{\partial f^{rs}_{p,t,i}} = \text{sd}^{rs}_{p,i}(t) + \sum_{a\in A_p}\text{LMC}^{rs}_{a,t,i}
\end{equation}
where $u_{a,i}(k)$ is the inflow of link $a$ for vehicle class $i$ at time $k$. Because we use one unit of perturbation flow to approximate PMCs and LMCs, $u_{a,i}(k)$ is set to be 1. $\omega_{a,i}(k,\textbf{f})$ is link travel time depending on overall network flow pattern $\textbf{f}$ and arrival time $k$. $k$ denotes the arrival times along successive links $a\in A_p$ where $A_p$ is the IDs of successive links along path $p$.}

\begin{proposition}
    The path-based SO-DTA can be transformed into an equivalent Variational Inequality Problem (VIP), which is to find $\bf{f}^{*} = \{f^{rs,*}_{p,t,i}\}_{rs,p,t,i}\in \varOmega$ such that
\begin{equation}
    \sum_{rs\in RS}\sum_{i\in I}\sum_{p\in P^{rs}_i}\int_{0}^T \text{PMC}^{rs}_{p,t,i}(\textbf{f}^{*}) \left(f^{rs}_{p,t,i} - f^{rs,*}_{p,t,i}\right)dt\geq 0,\ \forall \bf{f}\in \varOmega
\end{equation}
where the PMC can be represented in a sub-gradient form as follows without loss of generalization
\begin{equation}
    \text{PMC}^{rs}_{p,t,i}(\textbf{f}) \in \left[{\text{PMC}^{rs}_{p,t,i}(\textbf{f})}^{-}, {\text{PMC}^{rs}_{p,t,i}(\textbf{f})}^{+}\right] = \left[\lim_{x\uparrow f^{rs}_{p,t,i}}\frac{\bf{TC}(\bf{\tilde{f}}) - \bf{TC}(\bf{f})}{x - f^{rs}_{p,t,i}}, \lim_{x\downarrow f^{rs}_{p,t,i}}\frac{\bf{TC}(\bf{\tilde{f}}) - \bf{TC}(\bf{f})}{x - f^{rs}_{p,t,i}}\right]
\end{equation}
\end{proposition}
{The non-differentiable issue in PMC computation occurs when ${\text{PMC}^{rs}_{p,t,i}(\textbf{f})}^{-}\neq {\text{PMC}^{rs}_{p,t,i}(\textbf{f})}^{+}$, indicating adding one unit path flow and removing one unit path flow have impacts of different magnitudes on the generalized cost of the system \citep{zhang2020path}.}

In this study, we solve the SO-DTA using numerical solution algorithm based on discretized time intervals, and the continuous VIP is reformulated in a discrete form:
\begin{equation}
    \text{Find}\ \textbf{f}^{*}\in \Omega\ \text{such that} \sum_{rs\in RS}\sum_{i\in I}\sum_{p\in P^{rs}_i}\sum_{t\in T_a}\text{PMC}^{rs}_{p,t,i}(\textbf{f}^{*}) \left(f^{rs}_{p,t,i} - f^{rs,*}_{p,t,i}\right)\geq 0,\ \forall \bf{f}\in \varOmega
\end{equation}

\subsection{{Revisiting Multi-class Cell Transmission Model by \cite{qian2017modeling}}}
\label{sec:multi_CTM}
{In this study, the multi-class Cell Transmission Model (CTM) introduced by \cite{qian2017modeling} is used to model link traffic flow dynamics with heterogeneous vehicle classes. In this section, we summarize the key modeling components used throughout the subsequent PMC derivations. We present the case of two vehicle classes for clarity, and the extension to multiple classes follows the same logic.}

\subsubsection{{Fundamental Diagram of Two Vehicle Classes}}
\begin{figure}[H]
    \centering
    \includegraphics[scale = 0.4]{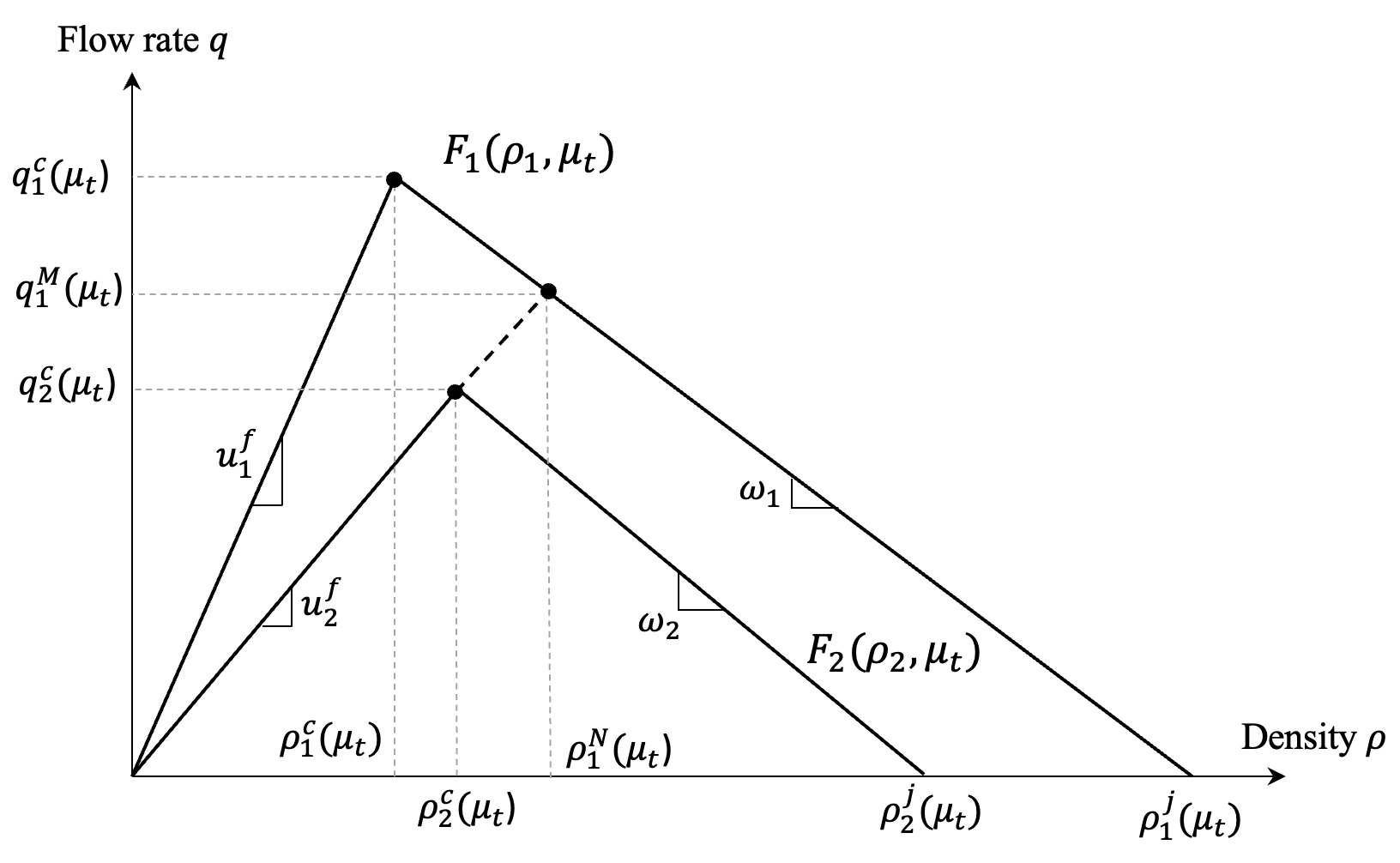}
    \caption{Fundamental diagram for the bi-class traffic stream}
    \label{fig:FD}
\end{figure}

{Suppose we have two vehicle classes, denoted by $i\in{1,2}$. Without loss of generality, class 2 has lower free-flow speed. 
Each vehicle class has identical car-following behavior and vehicle size, and has its respective homogeneous fundamental diagram (FD) $F_i(\rho_i)$, which governs the relationship between flow rate and density in homogeneous traffic conditions with class-$i$ vehicles. Recent studies found that link fundamental diagrams can vary with localized road configurations (e.g., curbside parking, lane blockage) \citep{koch2022physics, liu2024modeling}. We therefore generalize the multi-class CTM to accommodate time/location dependent road configurations. Let $\mu_t$ denote the local road configuration state at time $t$. The class-specific FD is written as $F_i(\rho_i,\mu_t),\ i\in\{1,2\}$, and its key parameters, including capacity $q_i^c(\mu_t)$, critical density $\rho_i^c(\mu_t)$ (at capacity) and jam density $\rho_i^j(\mu_t)$, may vary with $\mu_t$. Figure~\ref{fig:FD} illustrates the triangular FDs of the two vehicle classes used in this study.}
\subsubsection{{Road space splits and perceived densities}}
{When multiple vehicle classes are mixed, the congestion experienced by one class is jointly determined by the presence of both classes. This inter-class impact is characterized using the concept of perceived equivalent density (or perceived density), which allows each class to interpret the mixed traffic state through its own lens. \cite{qian2017modeling} introduce the road space split described by parameters $\alpha_1$ and $\alpha_2$, which represent the effective portion of road space associated with each class. Using class 1 as an example, class-1 vehicles perceive their own density $\rho_1$, and additionally perceive an equivalent density contributed by class 2, denoted as $\rho_2\,\delta_1(\alpha_1,\alpha_2)$, where $\delta_1(\cdot)$ is a conversion factor governed by the space split. The physical meaning is that, while the physical densities $(\rho_1,\rho_2)$ describe the actual mixed state, a class-1 vehicle perceives class-2 vehicles as if they were converted into class-1 vehicles occupying a certain effective share of the road space. Symmetrically, class 2 perceives its own density $\rho_2$ and an equivalent density $\rho_1\,\delta_2(\alpha_1,\alpha_2)$ contributed by class 1.} 

{To summarize, given actual densities $(\rho_1,\rho_2)$ on a road section, $\alpha_1$ and $\alpha_2$ determine the perceived equivalent densities $(\rho_1^p,\rho_2^p)$ for each class, then the cross-class lateral interaction can be represented. In \cite{qian2017modeling}, three possible regimes of cross-class interaction are discussed when $(\rho^p_1,\rho^p_2)$ and  $(\alpha_1^p,\alpha_2^p)$ satisfy different conditions:}

{\textbf{Free-flow regime.} Both classes move in free flow and their interaction is minimal. Let $\rho_i^c(\mu_t)$ be the critical density for class $i$ under road configuration state $\mu_t$. The minimum space fraction needed for class $i$ to maintain free flow is $\rho_i/\rho_i^c(\mu_t)$, meaning that
\begin{equation}
    \alpha_1 = \frac{\rho_1}{\rho^c_1(\mu_t)},\quad \alpha_2 = \frac{\rho_2}{\rho_2^c(\mu_t)}.
\end{equation}
The corresponding perceived densities are
\begin{equation}
    \rho^p_1 = \rho_1 + \rho_2\frac{\rho^c_1(\mu_t)}{\rho_2^c(\mu_t)},\quad
    \rho^p_2 = \rho_2 + \rho_1\frac{\rho^c_2(\mu_t)}{\rho_1^c(\mu_t)}.
\end{equation}
The condition of the free-flow regime is
\begin{equation}
    \rho^p_1\leq \rho^c_1(\mu_t),\quad \rho^p_2\leq \rho^c_2(\mu_t),
\end{equation}
which is equivalent to
\begin{equation}
    \alpha_1 + \alpha_2 \leq 1.
\end{equation}}

{\textbf{Semi-congested regime.} Class 1 (with larger free-flow speed) does not have sufficient space to remain in free flow and becomes mildly congested, while class 2 can still remain in free flow. For class-2 vehicles, they keep on free flow under $\mu_t$, yielding that
\begin{equation}
    \alpha_2 = \frac{\rho_2}{\rho_2^c(\mu_t)},\quad \alpha_1 = 1-\alpha_2.
\end{equation}
The perceived densities satisfy
\begin{equation}
    \rho^p_1 = \frac{\rho_1}{\alpha_1}\leq \rho^N_1,\quad
    \rho^p_2 = \frac{\rho_2}{\alpha_2} = \rho^c_2(\mu_t),
\end{equation}
where $\rho_1^N$ denotes the density threshold separating semi-congested and fully congested regimes, as shown in Figure~\ref{fig:FD}.}

{\textbf{Fully congested regime.} Both classes are congested and travel at the same speed because in the congestion traffic, class 1 cannot overtake class 2. We have
\begin{equation}
    u = \frac{F_1(\rho^p_1, \mu_t)}{\rho^p_1} = \frac{F_2(\rho^p_2, \mu_t)}{\rho^p_2}.
\end{equation}
Assuming triangular FDs (Fig.~\ref{fig:FD}), the space splits can be derived as
\begin{equation}
    \alpha_1 = \frac{w_1 - w_2 + \rho^j_2(\mu_t)
    \frac{w_2}{\rho_2}}{\rho^j_2(\mu_t)\frac{w_2}{\rho_2} + \rho^j_1(\mu_t)\frac{w_1}{\rho_1}},\quad
    \alpha_2 = \frac{w_2 - w_1 + \rho^j_1(\mu_t)
    \frac{w_1}{\rho_1}}{\rho^j_2(\mu_t)\frac{w_2}{\rho_2} + \rho^j_1(\mu_t)\frac{w_1}{\rho_1}}.
\end{equation}
Therefore, the perceived densities for fully congestion are
\begin{equation}
    \rho^p_1 = \frac{\rho_1}{\alpha_1} > \rho^N_1,\quad
    \rho^p_2 = \frac{\rho_2}{\alpha_2} > \rho^c_2(\mu_t).
\end{equation}
To summarize, the state of class-$i$ traffic depends on the corresponding perceived density that includes actual class-$i$ vehicles and converted class-$i$ vehicles from the other classes,
\begin{equation}
    \begin{aligned}
        F^{\text{mix}}_1(\rho_1, \rho_2, \mu_t) &= F_1(\rho_1^p, \mu_t)\\
        F^{\text{mix}}_2(\rho_1, \rho_2, \mu_t) &= F_2(\rho_2^p, \mu_t)\\
    \end{aligned}
\end{equation}
Note that on the boundary between the free-flow and semi-congested regimes, the perceived densities can equal the critical densities, i.e., the perceived inflows exactly match capacities. This boundary induces non-differentiability when deriving PMCs in the following sections.}

\subsubsection{{Flux between cells}}
{In the multi-class CTM, each link is discretized into cells of length $\Delta x$ and simulated with time step $\Delta t$, satisfying $\Delta x \geq u^f_1\Delta t$. Heterogeneous traffic propagates between adjacent cells using the supply-demand relations, where cross-class interaction is captured through the perceived densities. Consider two adjacent cells, an upstream cell with superscript $l$, and a downstream cell with superscript $r$. For simplicity, we omit the time index $t$ and road configuration $\mu_t$ in the following equations. The bi-class fluxes $q_1$ and $q_2$ between cells depend on the states of two adjacent cells, parameterized by $(\rho^{p,l}_i, \rho^{p,r}_i)\ i\in \{1,2\}$.}

{The demand of the left cell in terms of class $i\in \{1,2\}$ is defined as
\begin{equation}
    D_i^l = F_{i}^{l,\text{demand}}\left(\rho^{p,l}_i\right) = \min\left(q^{c,l}_i,\, u^f_i \rho^{p,l}_i \right)
\end{equation}
and the supply of the right cell in terms of class $i$ is defined as
\begin{equation}
    S_i^r = F_{i}^{r,\text{supply}}\!\left(\rho^{p,r}_i\right) = \min\left(q^{c,r}_i,\, w_i\left(\rho^{j,r}_i - \rho^{p,r}_i\right)\right)
\end{equation}
Note that the adjacent cells can be heterogeneous, meaning that $q^{c,l}_i$ and $q^{c,r}_i$ can be different based on specific cell conditions. The flux of class $i$ is specified as
\begin{equation}\label{eq:flux}
    q_i = \alpha_i^l \min\!\left(D_i^l,\, S_i^r\right)
\end{equation}
In \cite{qian2017modeling}, it is proved that the multiclass flux model in Equation~\ref{eq:flux} is consistent with the single-class CTM.}

{To summarize, at each time step in the network loading, for each cell on a link, we first obtain the physical traffic densities $(\rho_1,\rho_2)$ from the conservation update, then determine the traffic regime using the perceived densities $(\rho_1^p,\rho_2^p)$ and road space split $(\alpha_1,\alpha_2)$. Next, we evaluate class-specific cell demand and supply through FD $F_i(\rho_i^p,\mu_t)$ and determine the actual inter-cell flux $q_i$ to update flows.}
\section{{Approximating PMC of Heterogeneous Traffic and Application in SO-DTA}}
\label{sec:model}
\subsection{{Decomposing PMC for bi-class traffic}}
First, we {decompose} the total derivative of the total travel cost with respect to the {$p$-th path flow of vehicle class $i\in I$} as follows
\begin{equation}
    \text{PMC}^{rs}_{p,t,i}(\textbf{f}) = \frac{\partial\text{TC}(\textbf{f}_i)}{\partial f^{rs}_{p,t,i}} + \sum_{j\in I_{\backslash i}}\frac{\partial\text{TC}(\textbf{f}_j)}{\partial f^{rs}_{p,t,i}}
\end{equation}
where the first term indicates the total cost of $i$-th vehicle class with respect to path flow $f^{rs}_{p,t,i}$ and the second term is the rest of total cost of the other vehicle classes ($j\in I_{\backslash i}$) with respect to the $p$-th path flow of class $i$. The second term depicts the interaction between traffic of different vehicle classes, which is the focus of our study. {In the following analysis, we only consider the case of two vehicle classes, denoted by subscript 1 and 2, and more general cases can be extended easily. The bi-class SO-DTA can be formulated as the VIP in Equation~\ref{vi_2}.}
\begin{equation}
\label{vi_2}
\small
    \begin{aligned}
        \text{Find}\ \textbf{f}^{*}_1, \textbf{f}^{*}_2&\in \Omega\\
        \text{such that} &\sum_{rs\in RS}\sum_{t\in T_a}\left\{\sum_{p\in P^{rs}_1}\left[\frac{\partial\text{TC}(\textbf{f}_1)}{\partial f^{rs}_{p,t,1}} + \frac{\partial\text{TC}(\textbf{f}_2)}{\partial f^{rs}_{p,t,1}}\right]\left(f^{rs}_{p,t,1} - f^{rs,*}_{p,t,1}\right) + \sum_{p\in P^{rs}_2}\left[\frac{\partial\text{TC}(\textbf{f}_2)}{\partial f^{rs}_{p,t,2}} + \frac{\partial\text{TC}(\textbf{f}_1)}{\partial f^{rs}_{p,t,2}}\right]\left(f^{rs}_{p,t,2} - f^{rs,*}_{p,t,2}\right)\right\}\geq 0\\
        & \textbf{f}_1, \textbf{f}_2\in \Omega = \left\{\textbf{f}_1,\textbf{f}_2|\Delta_1 \textbf{f}_1 = \textbf{q}_1, \Delta_2 \textbf{f}_2 = \textbf{q}_2\right\} \\
    \end{aligned}
\end{equation}
In Equation~\ref{vi_2}, there are four PMC terms to be analyzed separately and each term has a corresponding sub-gradient with upper and lower bounds. For simplicity, we use $\text{PMC}^{j}_{i,t}$ indicates the derivatives of total cost of $j$-class path flow $\bf{f}_j$ over $i$-class path flow ${f}_{i,t}$ at time $t$, and we omit other subscripts for clarity. Similarly, link marginal costs are denoted by $\text{LMC}^j_{i,t}$.
\begin{equation}
    \text{PMC}^j_{i,t} = \frac{\partial \text{TC}(\bf{f}_j)}{\partial f_{i,t}}
\end{equation}
{The four PMC terms can be grouped into two categories: (1) intra-class PMCs which capture the marginal cost imposed by class $i\in\{1,2\}$ on itself (i.e., $\text{PMC}^1_{1,t}$ and $\text{PMC}^2_{2,t}$), and (2) inter-class PMCs which capture the marginal cost imposed by one class on the other due to cross-class interactions (i.e., $\text{PMC}^1_{2,t}$ and $\text{PMC}^2_{1,t}$).}

\subsubsection{Intra-class terms}
{We first analyze intra-class terms, $\text{PMC}^1_{1,t}$ and $\text{PMC}^2_{2,t}$. After each DNL run, inflow and outflow profiles are recorded in class-specific cumulative curves \cite{ma2020estimating, liu2024modeling}. We follow the work of \cite{qian2012system, zhang2020path} to analyze these class-specific PMCs but the main difference in our setting is how link states (i.e., regimes) are identified when traffic is heterogeneous, since the regime depends on class interactions and class-dependent parameters rather than a single homogeneous fundamental diagram.}

\begin{figure}[H]
    \centering
    \includegraphics[scale = 0.5]{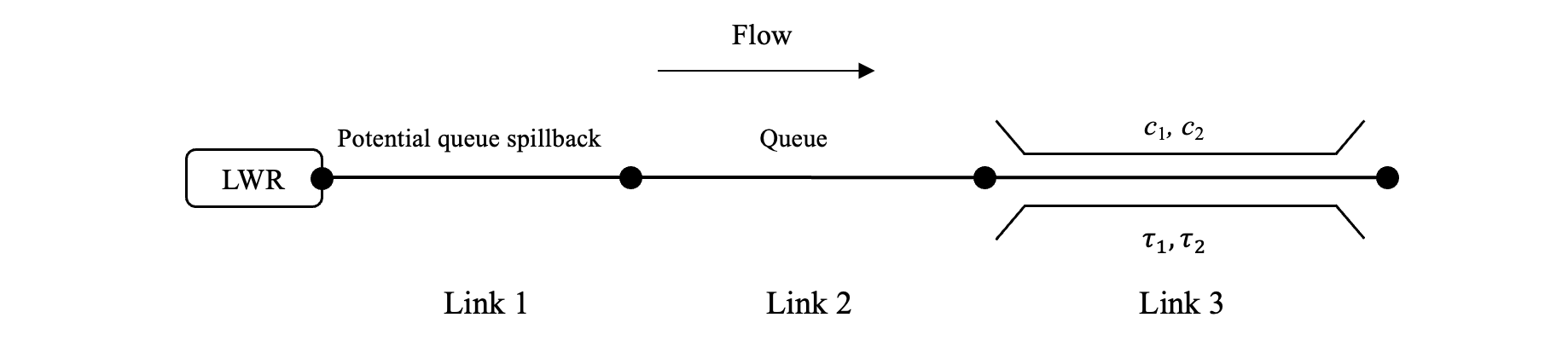}
    \caption{Single Bottleneck}
    \label{fig:single_bottleneck}
\end{figure}

Consider a simple network with three links shown in Figure~\ref{fig:single_bottleneck}. The bi-class traffic flow goes through the network from left to right. The Lighthill-Whitham-Richards (LWR) model is used as traffic flow dynamic model in the DNL. Without loss of generality, suppose link 3 is a bottleneck with capacity $c_1$ and $c_2$ for two separate classes and the free flow travel time for two classes on link 3 are $\tau_1$ and $\tau_2$. If the inflow exceeds the capacities, queues (of different classes) are accumulated on link 2 and may spillover upstream to link 1. This network has two path flows for two vehicle classes respectively, denoted by $f_1$ and $f_2$. In the multi-class DNL, space split ratios (i.e., $\alpha_1$ and $\alpha_2$) and the perceived densities (i.e., $\rho^p_{1,t}$ and $\rho^p_{2,t}$) are used to determine whether the bottleneck link 3 is active. {We first introduce some definitions that will be used in the following discussion.}

\begin{definition}
    \textbf{Active bottleneck}. In the case of multi-class traffic, a bottleneck is active at time $t$ for vehicle class $i$ when the perceived density of class $i$ at time $t$ is larger that its critical density, i.e., $\rho^p_{i,t} \geq \rho^c_{i}$. In bi-class case, the space split ratio of class $i$ should fall into the regime of either semi-congested or fully congested cases.
\end{definition}

\begin{definition}
    \textbf{Perturbation arrival time}. The perturbation arrival time is defined as the arrival time of the perturbation path flow on one specific link along the path. The perturbation arrival time governs whether LMC of current link is differentiable and depends on conditions of upstream links.
\end{definition}

\begin{definition}
    \textbf{Queue dissipation time}. The queue dissipation time for vehicle class $i$ is defined as the start time when the regime switches from semi-congested or fully congested back to free flow and the inflow equals to outflow.
\end{definition}

\begin{figure}[H]
    \centering
    \includegraphics[width=\linewidth]{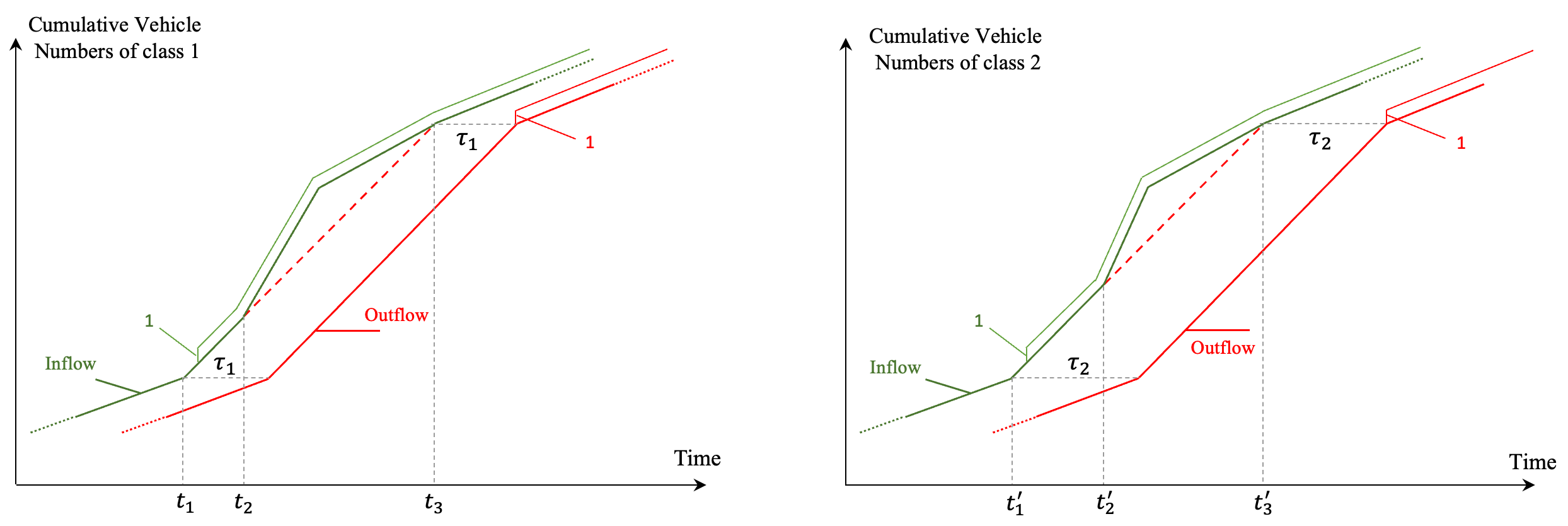}
    \caption{{Cumulative curves of two vehicle classes}}
    \label{fig:single}
\end{figure}

{Consider time-varying in/outflow profiles of two vehicle classes shown in Figure~\ref{fig:single}.} We assume no queue spillover occurs on link 1 and queue accumulates on link 2 due to the {potential} bottleneck ahead on link 3. Figure~\ref{fig:single} depicts the cumulative curves of two vehicle classes on the bottleneck separately. Before some time early than $t_1$ and $t_1^{'}$, traffic streams of both vehicle classes maintain free flow state and no inter-class impact needs to be considered. During this time, the bottleneck is inactive for both vehicle classes, which is the case that $\alpha_1$ and $\alpha_2$ fall into the free-flow regime and satisfies $\alpha_1 + \alpha_2 \leq 1$. Also the perceived densities are both lower than the critical densities, that is $\rho^p_{1,t} \leq \rho_1^c$ and $\rho^p_{2,t} \leq \rho_2^c$. In this case, $\text{PMC}^{1,+}_{1,t} = \text{PMC}^{1,-}_{1,t} = \tau_1$ and $\text{PMC}^{2,+}_{2,t} = \text{PMC}^{2,-}_{2,t} = \tau_2$ since any perturbation vehicle arriving the inactive bottleneck link will travel at its free flow speed, adding or removing one perturbation vehicle unit will increase or decrease the path travel time by the free flow travel time of the corresponding vehicle class.

{As the bi-class arrival rates increase with time}, traffic stream of class 1 might first reach the capacity at time $t_1$, meaning that the perceived density of class 1 equals to the critical density ($\rho^p_{1,t} = \rho^c_1$), while stream of class 2 is able to remain free flow. We assume from time $t_1$ to time $t_2$, $\rho^p_{1,t} = \rho^c_1$ and after $t_2$, the arrival rate exceeds the departure rate and queue forms. At time $t_3$, the queue of vehicle class 1 dissipates. Similarly, for traffic stream of class 2, there is a starting time $t_1^{'}$ (might be later than $t_1$) from when the perceived density $\rho^p_{2,t} = \rho_2^c$ and remains until $t_2^{'}$. After $t_2^{'}$ queue of vehicle class 2 forms and dissipate at time $t_3^{'}$. Starting at time $t_1 (t_1^{'})$, the bottlenecks for two vehicle classes becomes active. 

If the perturbation flow of vehicle class 1(2) arrives at the bottleneck at time $t > t_2 (t_2^{'})$ when there has been a queue accumulated, positive and negative flow perturbation will shift the departure curve after queue vanishes and the total travel time is differentiable. {However, according to the analysis in \cite{zhang2020path}, when the inflow rate first reaches the flow capacity (i.e., between $t_1$ and $t_2$), the path travel time is non-differentiable and the following proposition holds for the bi-class traffic setting.}

\begin{proposition}\label{prop2}
    The travel time is non-differentiable for class $i$ when the following conditions are met: (1) a downstream bottleneck is active for class $i$ along the path and (2) the perturbation arrival time $t$ at the bottleneck link satisfies that $\rho^p_{i,t} = \rho^c_{i}$ and the $\alpha_i$ falls into the regime of semi-congested case.
\end{proposition}

For vehicle class 1, the perturbation arrives between time $t_1$ and $t_2$ and $\rho^p_{1,t} = \rho_1^c$, meaning that the perceived flow of vehicle 1 equals exactly to the capacity $c_1$. After $t_2$ queue accumulated at the bottleneck. From some time after $t_2$, the perceived inflow rate decreases and the queue starts to vanish when inflow is lower than capacity $c_1$. Finally the queue diminishes at time $t_3$. When the flow perturbation arrives at the bottleneck at time $t\in [t_2, t_3]$, $\text{PMC}^{1}_{1,t} = t_3 - t + \tau_1$. When the flow perturbation arrives at the bottleneck at time $t\in [t_1, t_2)$, the path travel time is not differentiable, meaning $\text{PMC}^{1,+}_{1,t} \neq \text{PMC}^{1,-}_{1,t}$. If one marginal unit of vehicle 1 was taken out of the link at time $t\in [t_1, t_2)$, $\text{PMC}^{1,-}_{1,t} = \tau_1$ and if one marginal unit is added to the link, $\text{PMC}^{1,+}_{1,t} = t_3 - t + \tau_1$.

To summarize
\begin{equation}\label{eq:lmc_1}
    \text{PMC}^{1}_{1,t} = \begin{cases}
        \tau_1 & t < t_1\\
        [\tau_1, t_3 - t + \tau_1] & t\in [t_1, t_2)\\
        t_3 - t + \tau_1 & t\in [t_2, t_3)\\
        \tau_1 & t \geq t_3\\
    \end{cases}
\end{equation}

Similarly, the $\text{PMC}^{2}_{2,t}$ can be derived based on the arrival time of perturbation unit at bottleneck and the times when perceived inflow reaches $c_2$ ($t\in [t_1^{'}, t_2^{'})$) and queue dissipation time $t_3^{'}$. The $\text{PMC}^{2}_{2,t}$ can be summarized as
\begin{equation}
    \text{PMC}^{2}_{2,t} = \begin{cases}
        \tau_2 & t < t_1^{'}\\
        \left[\tau_2, t_3^{'} - t + \tau_2\right] & t\in [t_1^{'}, t_2^{'})\\
        t_3^{'} - t + \tau_2 & t\in [t_2^{'}, t_3^{'})\\
        \tau_2 & t \geq t_3^{'}\\
    \end{cases}
\end{equation}

\subsubsection{Inter-class terms}
{When interactions between different vehicle classes are considered, we adopt a pragmatic approach to approximate the inter-class PMCs using intra-class PMCs and the corresponding conversion factors. We use the link-state regimes introduced in \cite{qian2017modeling} because our multi-class DNL is rooted in this framework, and it is therefore consistent to use the same set of parameters to model cross-class interactions. Other approaches can be adopted as long as they are compatible with the DNL in use. The key intuition is to derive an equivalent conversion factor based on the link regime. Specifically, we compute the inter-class terms by scaling the intra-class terms,}
\begin{equation}
    \text{PMC}^1_{2,t} = \delta_2^1 \cdot \text{PMC}^1_{1,t} \quad \text{and}\quad \text{PMC}^2_{1,t} = {\delta_1^2} \cdot \text{PMC}^2_{2,t}
\end{equation}
where $\delta_2^1$ indicates the conversion factors depending on the road space split ratios. Consistently, we define them as,
\begin{equation}
\begin{aligned}
        \delta_2^1(\alpha_1, \alpha_2) &= \frac{\rho_1/\alpha_1}{\rho_2/\alpha_2}\\
        \delta_1^2(\alpha_1, \alpha_2) &= \frac{\rho_2/\alpha_2}{\rho_1/\alpha_1}\\       
\end{aligned}
\end{equation}
where both $\rho_1$ and $\rho_2$ are both nonzero. As discussed in Section~\ref{sec:multi_CTM}, three regimes are determined by perceived densities and road space splits. We discuss each regime separately to derive the inter-class PMC terms.
When the link state falls into the free-flow regime, the interaction between the two classes can be neglected, meaning that adding one unit of flow (either car or truck) does not influence the other class. Therefore, the inter-class PMCs are
{\begin{equation}
\label{eq:regime1}
\begin{aligned}
        \text{PMC}^2_{1,t} &= 0\\
        \text{PMC}^1_{2,t} &= 0
\end{aligned}
\end{equation}}
In the semi-congested regime, we consider one-way interaction, which is adding one unit of truck flow affect cars, but car flow does not influence trucks because trucks can maintain free flow speed. In this case, we have
\begin{equation}
\label{eq:regime2}
\begin{aligned}
        \text{PMC}^2_{1,t} &= \delta_1^2(\alpha_1, \alpha_2)\cdot \text{PMC}^2_{2,t} = \frac{\rho^p_2 - \rho_2}{\rho_1}\cdot \text{PMC}^2_{2,t}\\
        \text{PMC}^1_{2,t} &= 0
\end{aligned}
\end{equation}
In the fully-congestion regime, the two classes interact with each other as they compete for limited road space and neither can maintain free-flow speeds, and the interaction terms can be computed by
\begin{equation}
\label{eq:regime3}
    \begin{aligned}
        \text{PMC}^2_{1,t} &= \delta_1^2(\alpha_1, \alpha_2)\cdot \text{PMC}^2_{2,t} = \frac{w_1\rho_2 + (\rho_2^j - \rho_2)w_2}{w_2\rho_1 + (\rho_1^j - \rho_1)w_1} \text{PMC}^2_{2,t}\\
        \text{PMC}^1_{2,t} &= \delta_2^1(\alpha_1, \alpha_2)\cdot \text{PMC}^1_{1,t} = \frac{w_2\rho_1 + (\rho_1^j - \rho_1)w_1}{w_1\rho_2 + (\rho_2^j - \rho_2)w_2}\text{PMC}^1_{1,t}
\end{aligned}
\end{equation}
{Note that when computing PMCs, the non-differentiable issue may not occur simultaneously for cars and trucks, which depends on the actual arrival time of the class-specific unit flow. Therefore, we should record the time-dependent link states during DNL and use this information to determine the appropriate PMC expressions.}

In summary, for a general network, the PMCs can be computed as follows: (1) for each time interval, use the time-varying perceived densities of the two vehicle classes to determine link conditions and if non-differentiable issue occurs according to Proposition~\ref{prop2}; (2) add one unit class-specific path flow and trace each link sequentially along the path, compute the time when the perturbation enters link $e$, denoted by $t^e_{i,\text{in}}$ and the time when the perturbation leaves link $e$, denoted by $t^e_{i,\text{out}}$; (3) calculate the LMC by
\begin{equation}
\begin{aligned}
    \text{LMC}_{i,e,t}^{i,+/-} &= t^{e,+/-}_{i,\text{out}} - t^{e,+/-}_{i,\text{in}} + \text{fft}_{i,e}\\
    \text{LMC}_{j,e,t}^{i,+/-} &= \delta^{i}_{j}\cdot \text{LMC}_{i,e,t}^{i,+/-}
\end{aligned}
\end{equation}
where $\text{fft}_{i,e}$ is the free flow travel time of class $i$ on link $e$, and the choice of $+/-$ depends on whether non-differentiable issue occurs when the perturbation arrives at current link. The PMC is obtained by aggregating LMCs along the path using Equation~\ref{eq:pmc_lmc}. 

{Here we also discuss a general approach for calculating PMCs under heterogeneous road configurations induced by on-street parking. As demonstrated in \cite{liu2024modeling}, curbside parking can have localized negative impact on link performance, leading to location-dependent within-link heterogeneity in the FD when a cell transmission model is used to describe link dynamics. Because the multi-class CTM naturally support cell heterogeneity, we can simply record cell-level inflow and outflow using more granular cumulative curves and compute the marginal cost at the cell level. The link-level LMC is then obtained by aggregating cell-level marginal costs along the link. The non-differentiability condition can also be checked cell-by-cell using the perceived densities of multi-class traffic. We leave this general extension to future research.}

\subsection{Solution Algorithm for PMC-based SO-DTA}
{In this section, we solve the discretized SO-DTA VIP formulation by a standard Method of Successive Averages (MSA) algorithm. The key idea is that at each iteration, we evaluate intra- and inter-class PMCs for each vehicle class under the current bi-class flow pattern and then move demand toward the path with the least PMC value. The resulting algorithm is heuristic but widely used in prior PMC-based SO-DTA studies.}

{Each MSA iteration consists of four main steps: (1) a bi-class DNL run to obtain link and path states; (2) PMC evaluation by tracing one unit perturbation path flow through the network; (3) a least PMC finding process based on the time dependent shortest path method for each OD pair and class; (4) the MSA averaging update. The algorithm is shown in Algorithm~\ref{alg}.}

{At the \textbf{Finding least PMC} step in Algorithm~\ref{alg}, we need to solve a time-dependent shortest path problem to find the least class-specific PMC problem. This can be solved by the decreasing order of time (DOT) algorithm introduced by \citep{chabini1998discrete}. More discussions under point queue model and LWR models can be found in \citep{qian2012system}. In this algorithm, we use a diminishing step size defined as
\begin{equation}
    \lambda^v = \frac{1}{1 + v}
\end{equation}}

{The convergence criteria in Algorithm~\ref{alg} is based on the gap function $g^\nu$ at iteration $\nu$, defined in Equation~\ref{eq:gap}.}
{\begin{equation}
    \label{eq:gap}
    g^\nu = \frac{\sum_{i}\sum_{t}\sum_{rs}\sum_{p}f^{rs}_{p,i}(t)(\text{PMC}^{rs}_{p,i}(t) - \mu^{rs}_{i})}{\sum_{i}\sum_{t}\sum_{rs}\sum_{p}f^{rs}_{p,i}(t)\mu^{rs}_{i}}
\end{equation}}
{where $\mu^{rs}_{i}$ denotes the minimum PMC for vehicle class $i$ and OD pair $rs$ over all feasible path and departure time combinations. Because we consider both route and departure time choice, $\mu^{rs}_{i}$ is constant with respect to $t$. If only route choice is considered, the corresponding term should be $\mu^{rs}_{i}(t)$, which depends on departure time $t$ as well.}

{When PMCs are non-differentiable, the upper/lower subgradient bounds differ. In the MSA algorithm, either lower bound or upper bound can be used or a convex combination of both bounds can be used. In this study we report results under the specified bound choice and leave advanced subgradient utilization to future work.}

\begin{algorithm}[H]
\caption{A vanilla MSA algorithm for SO-DTA with simultaneously route and departure time choices}
\label{alg}
{\begin{algorithmic}[1]
\State \textbf{Initialization:} any $\mathbf{f}_i^0 \in \Omega$ where $\Omega$ is the constraints of multi-class demand; $\nu = 0$ ; $\lambda^0$;
\Repeat
    \State \textbf{Bi-class DNL}. Load multi-class path flow $\mathbf{f}_i^\nu$ into the network;
    \State \textbf{Computing intra-class terms}. Tracking perturbation unit flow along each path and compute intra-class LMCs for all links based on Equation~\ref{eq:lmc_1}; 
    \State \textbf{Computing inter-class terms}. Determine the link regime (free-flow / semi-congested / fully congested) from the multi-class link states, and compute inter-class terms using the regime-dependent conversion factors using Equation~\ref{eq:regime2} and \ref{eq:regime3}; 
    \State \textbf{Aggregate LMCs to PMCs}. Aggregate LMCs over links on each path and include scheduled delay term to obtain PMCs using Equation~\ref{eq:pmc_lmc};
    \State \textbf{Least PMC search}. For each $rs \in RS$ ($rs \in RS$ given any $t \in T_d$), find the time-dependent path $[p_i^*, t^*]$ with least PMC separately for vehicle class $i$, which is to solve problem $[p_i^*, t^*] = \arg\min_{p,t}\text{PMC}^{rs}_{p,t,i}$ for each class. This is implemented as a time-dependent shortest path problem (e.g., DOT algorithm);
    \State \textbf{Auxiliary path flow}. Generate an auxiliary path flow pattern $\mathbf{g}_i(\mathbf{f}_i^\nu)$ by assigning all class-specific demand of $Q^{rs}_i$ onto $[p_i^*, t^*]$;
    \State \textbf{MSA update} $\mathbf{f}_i^{\nu+1} = (1 - \lambda^\nu)\mathbf{f}_i^\nu + \lambda^\nu \mathbf{g}_i(\mathbf{f}_i^\nu)$;
    \State \textbf{Step size update} $\lambda^\nu$ to $\lambda^{\nu+1}$, $\nu = \nu + 1$;
\Until{Convergence criteria meets;}
\end{algorithmic}}
\end{algorithm}

\section{Numerical Experiments}
\label{sec:experiment}
\subsection{A small network}
\begin{figure}
    \centering
    \includegraphics[width=\linewidth]{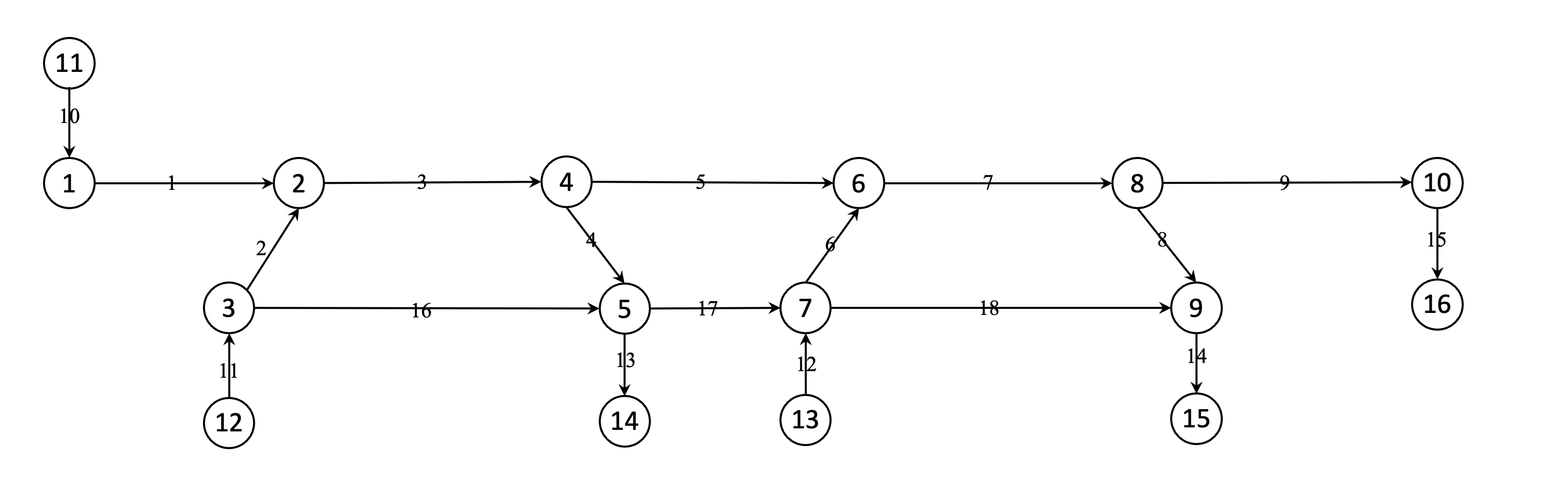}
    \caption{Small network}
    \label{fig:nie}
\end{figure}

{We first use a small corridor network which was used by \cite{zhang2020path}, shown in Figure~\ref{fig:nie}.} The link attributes are summarized in Table~\ref{tab:nie}. {The links with IDs 10-15 are OD connectors that are modeled using Point Queue with infinite large capacity and jam density values (set to be large constant in our simulation) and no congestion occurs on them.The other links represent road segments with traffic flow characteristics aggregated by lanes.}
The assignment time interval is 15 minutes, which determines how frequent path costs are updated and time-varying path choice proportion are calculated and used to assign demand. The loading time interval is 5 seconds. The whole DNL process has 10 assignment time intervals, which is totally 2.5 hours, we assume the realistic time is from 8:00 AM to 10:30 AM. 
We consider both route and departure time choices and set the travel cost parameters as: $\alpha = 1$, $\beta = 0.5$ and $\gamma = 2$. The punctual time window is from 8:30 AM to 10:00 AM.

{We report total travel cost (TTC) and total travel time cost (TTTC) to evaluate model performance. TTC is computed following Equation~\ref{eq:ttc}. In our experiments, we consider both route and departure time choice, so the generalized travel cost for each path flow has two components: (i) travel-time cost (path travel time times value of time) and (ii) schedule-delay penalty determined by the difference between the actual arrival time and the desired arrival time window at destinations. We compute TTC using path flows at the assignment time interval resolution (i.e., 15 mins). Consider the $p$-th path flow of vehicle class $i$ departing at time interval $t$ between OD pair $rs$, denoted by $f^{rs}_{p,i}(t)$, we extract the corresponding path travel time $w^{rs}_{p,i}(t)$ (in secs) from departure/arrival cumulative curves of the successive links along path $p$. The actual arrival time of this path flow is $t_{arr} = t + w^{rs}_{p,i}(t)$, represented in seconds. Given the experiment setting, the punctual arrival time window is 8:30-10:00 AM, which yields the target arrival time $t^*_{rs,i} = 4500$ (seconds) and tolerance term $\delta = 2700$ (seconds). The schedule delay cost for this path flow is determined by which region $\Delta t = t_{arr} - t^*_{rs,i}$ falls into as specified in Equation~\ref{eq:schedule}. We multiplied the penalty by path flow and then sum over all $(rs, p, t, i)$ to obtain the total schedule delay cost (TSDC)
\begin{equation}
    \text{TSDC} = \sum_{rs}\sum_{i}\sum_{t}\sum_{p} f^{rs}_{p,i}(t)\cdot \text{sd}(t + w^{rs}_{p,i}(t) -  t^*_{rs,i})
\end{equation}
Similarly, the TTTC is aggregated as
\begin{equation}
    \text{TTTC} = \sum_{rs}\sum_{i}\sum_{t}\sum_{p} f^{rs}_{p,i}(t)\cdot \alpha w^{rs}_{p,i}(t)
\end{equation}
and overall TTC is
\begin{equation}
    \text{TTC} = \text{TTTC} + \text{TSDC}
\end{equation}}

\begin{table}
    \scriptsize
	\caption{{Link properties of the small network (traffic flow characteristics are aggregated by lane)}}
	\label{tab:nie}
	\centering
    \begin{tabular}{ccccccccc}
\hline
        ID & length & lane & car free-flow speed & truck free-flow speed & car flow capacity & truck flow capacity & car jam density & truck jam density \\
         & (mile) & number & (mile/h) & (mile/h) & (veh/h) & (veh/h) & (veh/mile) & (veh/mile) \\
\hline
        1 & 1 & 3 & 50 & 40 & 6000 & 3600 & 540 & 240\\
        2 & 0.5 & 1 & 50 & 40 & 2000 & 1200 & 180 & 80\\
        3 & 2 & 3 & 50 & 40 & 6000 & 3600 & 540 & 240\\
        4 & 0.5 & 1 & 50 & 40 & 2000 & 1200 & 180 & 80\\
        5 & 2 & 3 & 50 & 40 & 6000 & 3600 & 540 & 240\\
        6 & 0.5 & 1 & 50 & 40 & 2000 & 1200 & 180 & 80\\
        7 & 1 & 2 & 50 & 40 & 4000 & 2400 & 360 & 160\\
        8 & 0.5 & 1 & 50 & 40 & 2000 & 1200 & 180 & 80\\
        9 & 1 & 3 & 50 & 40 & 6000 & 3600 & 540 & 240\\
        16 & 2 & 1 & 30 & 20 & 2000 & 1200 & 180 & 80\\
        17 & 2 & 1 & 30 & 20 & 2000 & 1200 & 180 & 80\\
        18 & 1 & 1 & 30 & 20 & 2000 & 1200 & 180 & 80\\
\hline
    \end{tabular}
\end{table}

We first solve SO-DTA using lower bounds of sub-gradients with and without considering inter-class PMC terms separately. The reason we choose lower bounds as they reflect the non-differentiability because when LMC is not differentiable, the lower bound is different from upper bound as discussed in previous sections. We use dynamic user equilibrium as the benchmark to evaluate the travel cost reductions. 

\begin{table}
\small
\centering
\begin{tabular}{lcccccc}
\hline
\textbf{Model Type} & \multicolumn{2}{c}{\textbf{Total Travel Time Cost (TTTC)}} & \multicolumn{2}{c}{\textbf{Total Travel Cost (TTC)}} & \multicolumn{2}{c}{\textbf{Gap}} \\
\cline{2-7}
& car        & truck      & car         & truck       & car         & truck       \\
\hline
DUE         & 978.2  & 191.1     &  2172.6    &   416.9    &    0.058    &     0.054    \\
DSO, L, intra + inter        & 724.5  & 147.8 &   1479.2   &   292.1    &    0.181    &   0.136    \\
DSO, L, intra only    & 762.3 & 151.1     &   1560.3   &   304.8    &   0.177   &   0.161    \\
DSO, U, intra + inter     & 852.9  & 164.2     &   1856.3   &   351.3  &  0.15     &    0.11    \\
DSO, U, intra only    & 856.5  & 165.1     &   1862.0   &    352.9   &   0.16    &    0.10   \\
\hline
\end{tabular}
\caption{Performance comparison of models}
\label{tab:due_dso}
\end{table}

Table~\ref{tab:due_dso} compares the performance of different traffic assignment models, including the benchmark of Dynamic User Equilibrium (DUE) and Dynamic System Optimum (DSO) with various settings. The DSO models are differentiated by whether they incorporate inter-class PMCs (intra + inter) or only use intra-class PMCs (intra only), and by whether they use the lower-bound (L) of PMCs or upper-bound (U) of PMCs in MSA updates. The travel costs are in unit of veh$\cdot$hour.

Results demonstrate the system-wide efficiency gains achieved by DSO models relative to the benchmark of DUE. For cars, the total travel time cost (TTTC) decreases from 978.2 under DUE to as low as 724.5 under DSO (L, intra + inter), representing a 26\% reduction. Trucks experience a similar improvement, with TTTC dropping from 191.1 to 147.8 (a 23\% reduction). Total travel cost (TTC), which incorporates both time and scheduled delay costs, follows the same pattern: cars reduce from 2172.6 to 1479.2 (32\% reduction), and trucks from 416.9 to 292.1 (30\% reduction). These results confirm that system-optimal traffic assignment can substantially improve overall network efficiency by considering inter-class marginal costs. 

Comparing the intra + inter and intra-only DSO models reveals the benefits of considering inter-class impacts and non-differentiable issues on traffic flow. For both lower- and upper-bound formulations, incorporating inter-class PMCs consistently yields lower TTTC and TTC. For instance, for cars under the lower-bound DSO, TTTC decreases from 762.3 (intra only) to 724.5 (intra + inter), while for trucks, TTTC decreases from 151.1 to 147.8. The TTC reductions show a similar pattern, highlighting that ignoring inter-class effects results in suboptimal allocations where cars and trucks compete for capacity inefficiently. Furthermore, the lower-bound DSO consistently achieves lower TTTC and TTC for both vehicle classes, regardless of whether inter-class PMCs are included. Because the difference between lower bound and upper bound incurs only when link capacity of mixed traffic flow reaches the capacity, the differences indicate that the lower-bound DSO with the consideration of non-differentiable issues can more effectively drive the network toward the system optimum.
\begin{figure}[H]
    \centering
    \includegraphics[width=0.7\linewidth]{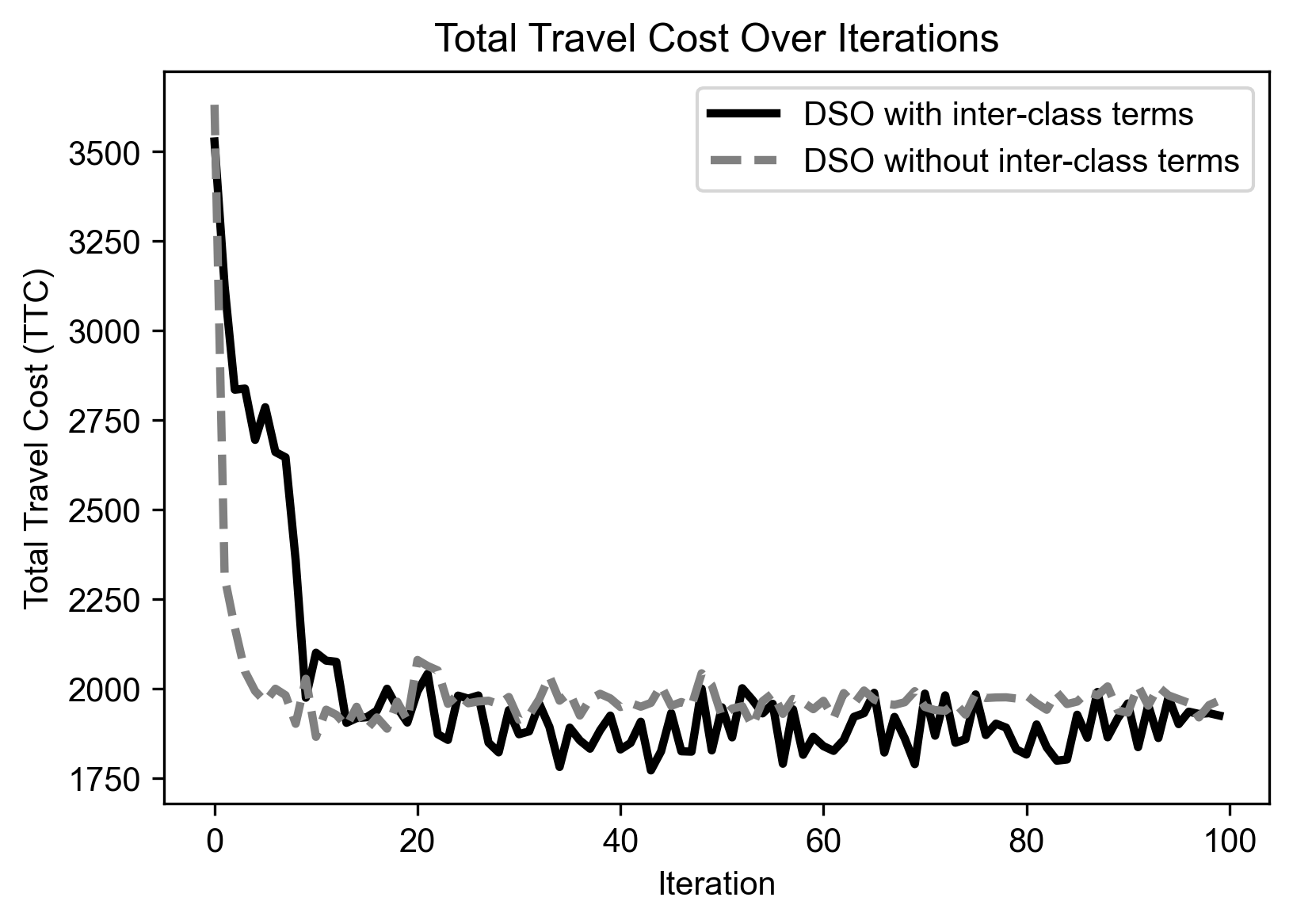}
    \caption{{Convergence of two DSO runs (with/without inter-class terms) on small network}}
    \label{fig:dso_convergence}
\end{figure}

\begin{figure}[H]
    \centering
    \includegraphics[width=0.8\linewidth]{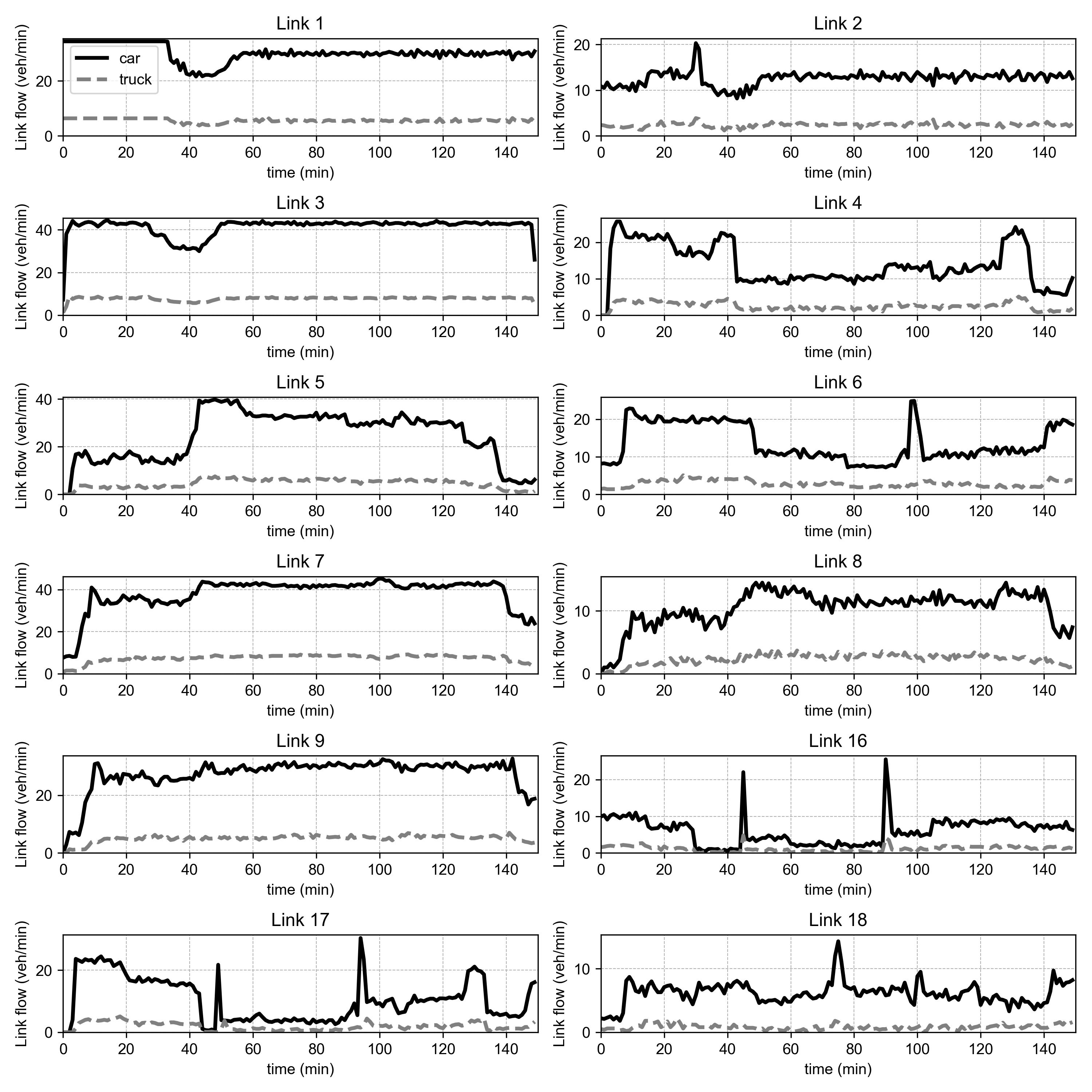}
    \caption{Link flow temporal pattern (DSO without inter-class PMCs)}
    \label{fig:link_inflow_dso_wo}
\end{figure}

{Figure~\ref{fig:dso_convergence} compares convergence behavior of two DSO runs. One is DSO with interaction (i.e., DSO, L, intra+inter in Table~\ref{tab:due_dso}) and the other is DSO without interaction (i.e., DSO, L, intra only in Table~\ref{tab:due_dso}). It can be observed that DSO without interaction converge earlier (after around 10 iteration) with larger total travel cost and DSO with interaction can further reduce total travel cost after 20 iteration and finally achieve a better DSO solution.}
\begin{figure}[H]
    \centering
    \includegraphics[width=0.8\linewidth]{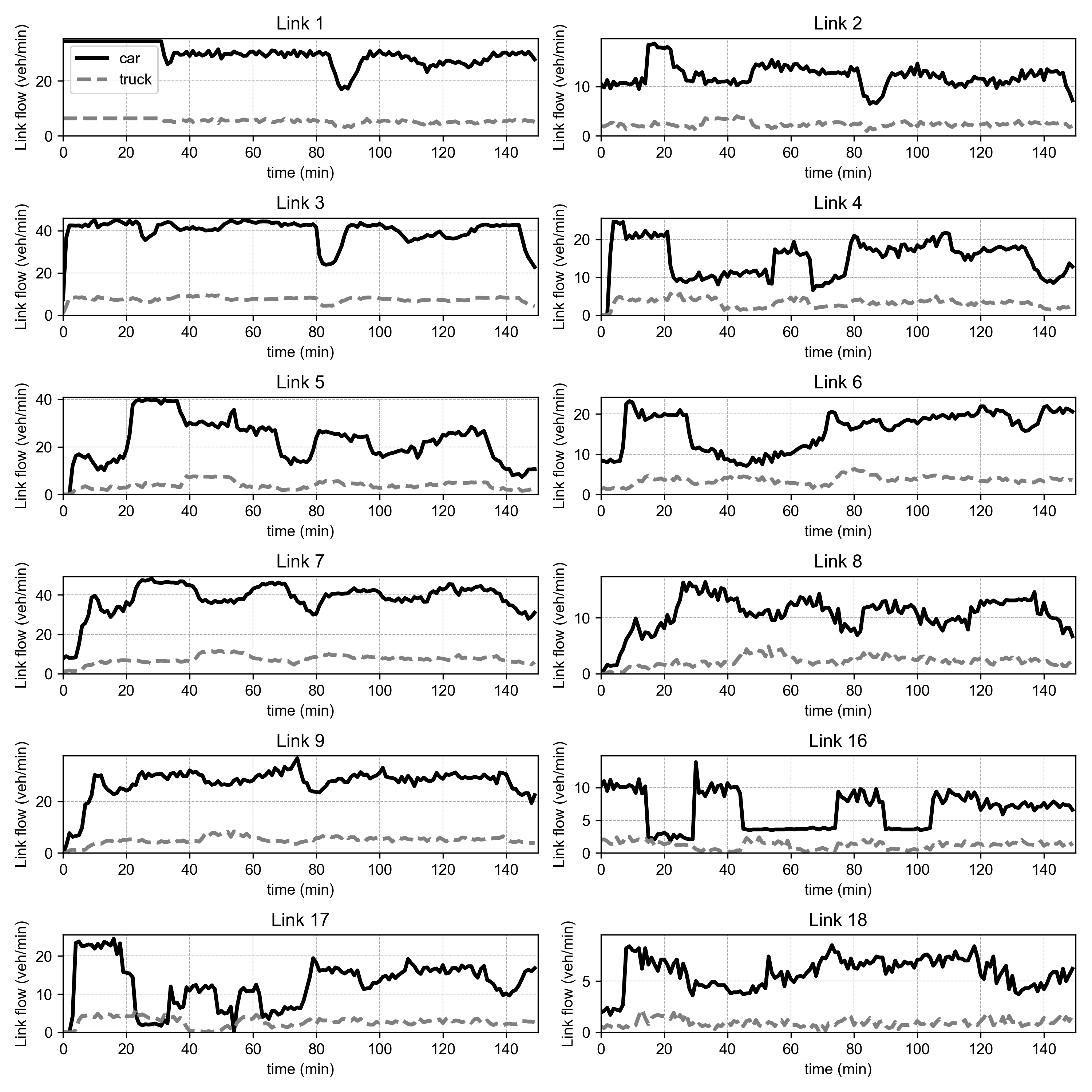}
    \caption{Link flow temporal pattern (DSO with inter-class PMCs)}
    \label{fig:link_inflow_dso_w}
\end{figure}
We further investigate the class-specific link flows {diversions in path and departure time choices} to compare link/path states at DSO with and without considering the bi-class interactions. Figures~\ref{fig:link_inflow_dso_w} and \ref{fig:link_inflow_dso_wo} compare the link-level flow patterns at DSO with and without multi-class interaction when computing PMCs in the MSA algorithm. 
It can be seen that the two runs exhibit different temporal patterns. When multi-class interaction is incorporated, both car and truck flows show smoother temporal variations without sharp spikes and a more balanced redistribution across links, indicating that the algorithm effectively captures cross-class congestion externalities. 
In contrast, the solution without multi-class interaction shows abrupt spikes and drops in both car and truck flows on several links (e.g., Links 2, 6, and 16), although some links have more stable temporal patterns (e.g., link 1, 3 and 7). This suggests that this DSO optimization treats each class independently and overlooks their mutual impacts on congestion, especially on local segments with limited capacity. 
Another noticeable pattern in {Figure~\ref{fig:link_inflow_dso_w}} is that, during certain time intervals {(e.g., 40–60 minutes on links 6, 8, and 9),} car and truck flows display inverse temporal trends: car flow decreases while truck flow increases, and vice versa. This pattern highlights the model’s ability to dynamically shift demand between classes to balance link utilization, demonstrating the potential of inter-class PMCs to coordinate multi-class flows rather than merely optimizing each class independently. The smoother, inversely correlated, and more evenly distributed flow patterns suggest that explicitly accounting for inter-class externalities {yields a better and more reasonable system optimum solution.}

\begin{figure}[H]
    \centering
    \includegraphics[width=\linewidth]{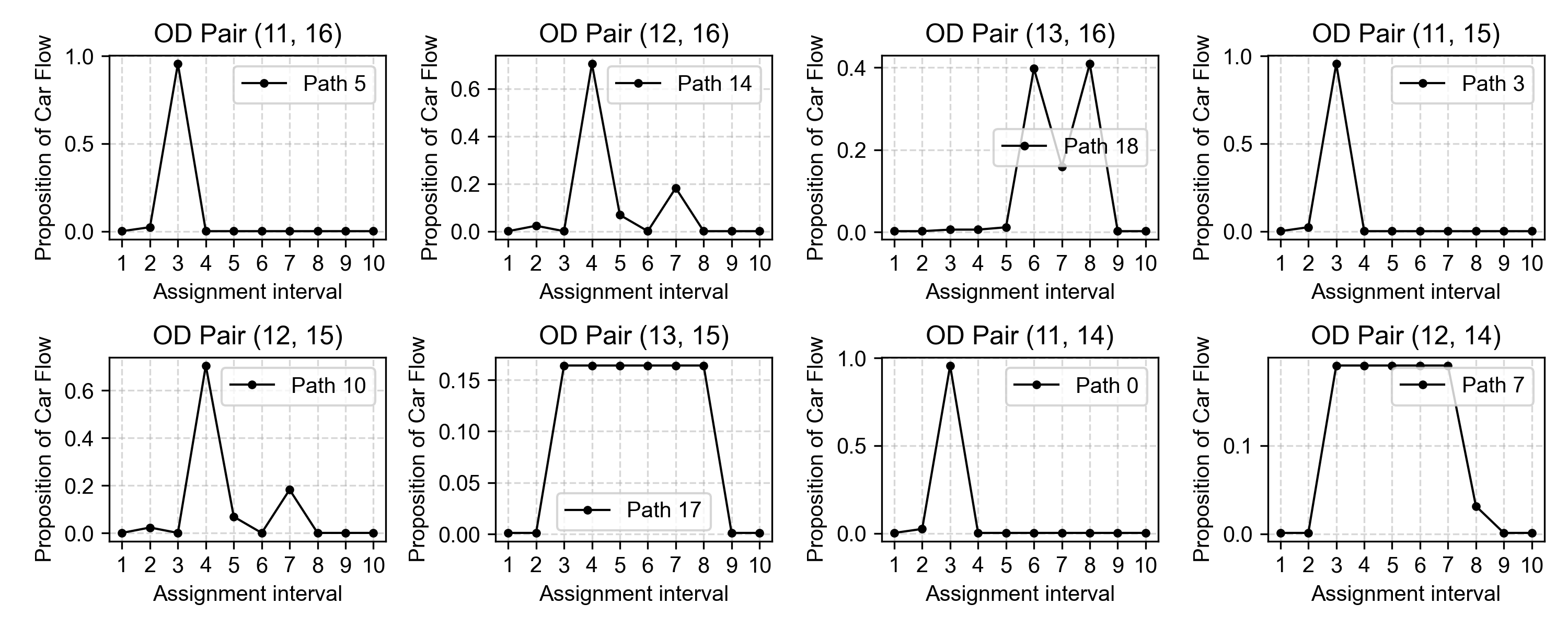}
    \caption{Car path flow diversion (with inter-class PMCs)}
    \label{fig:od_diversion_dso_w_car}
\end{figure}

\begin{figure}[H]
    \centering
    \includegraphics[width=\linewidth]{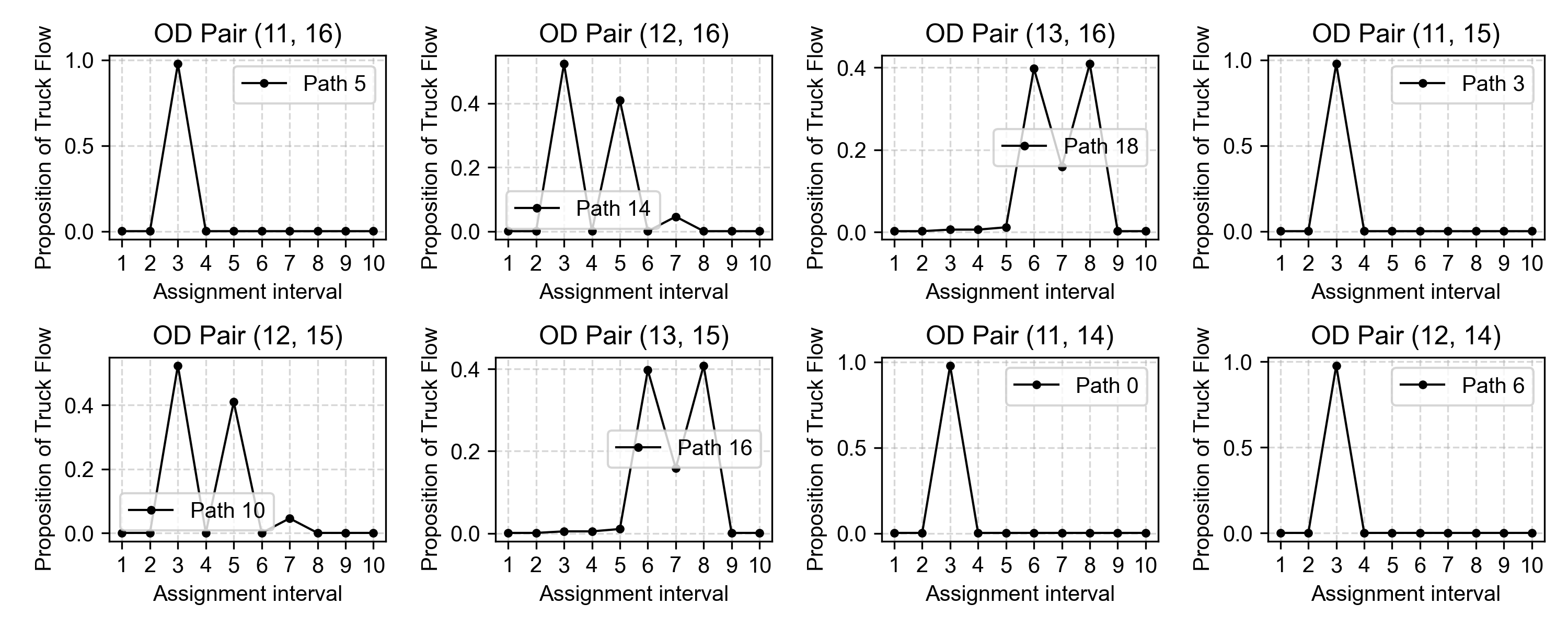}
    \caption{Truck path flow diversion (with inter-class PMCs)}
    \label{fig:od_diversion_dso_w_truck}
\end{figure}

Figures~\ref{fig:od_diversion_dso_w_car} and \ref{fig:od_diversion_dso_w_truck} show the joint path and departure time choice for cars and trucks, respectively, at DSO with inter-class PMCs. At convergence, each OD pair ultimately assigns flow to a single dominant path, and the other paths are not plotted because their proportions are close to zeros. 
A clear temporal difference between vehicle classes can be observed for some OD pairs (e.g., pairs (12, 16), (12, 15), (13, 15) and (12, 14)), where cars and trucks exhibit distinct but complementary temporal path flows. This indicates that inter-class PMCs effectively redistribute multi-class demand {across paths and departure times, mitigating the cross-class congestion}.

In contrast, other OD pairs (e.g., (11, 16), (11, 14), (11, 15) and (13, 16)) display strong alignment in path and departure time choices across vehicle classes, implying that for these OD pairs, {inter-class effects are relatively weak and insufficient to induce spatio-temporal separation of multi-class flows. The intra-class PMCs dominate the choice decisions, resulting to both classes converging to similar optimal path and departure time choice.}

\begin{figure}[H]
    \centering
    \includegraphics[width=\linewidth]{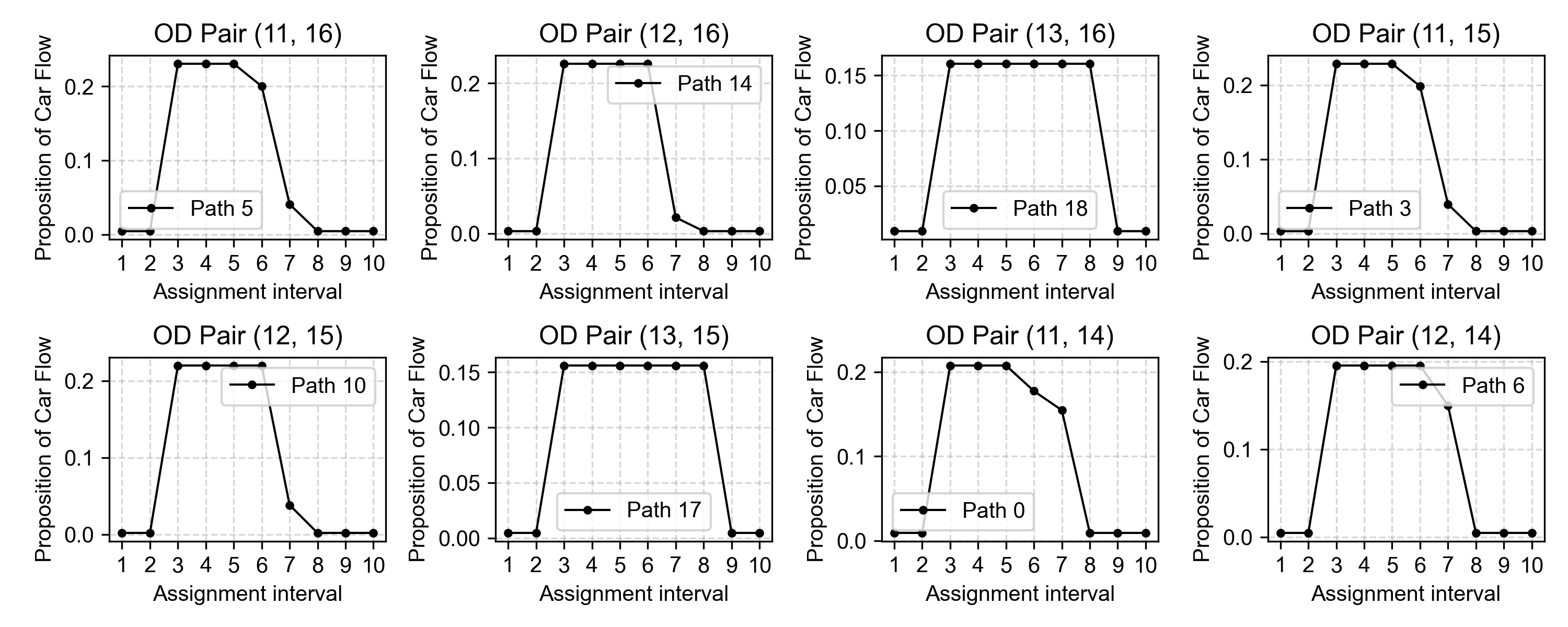}
    \caption{Car path flow diversion (without inter-class PMCs)}
    \label{fig:od_diversion_dso_wo_car}
\end{figure}

\begin{figure}[H]
    \centering
    \includegraphics[width=\linewidth]{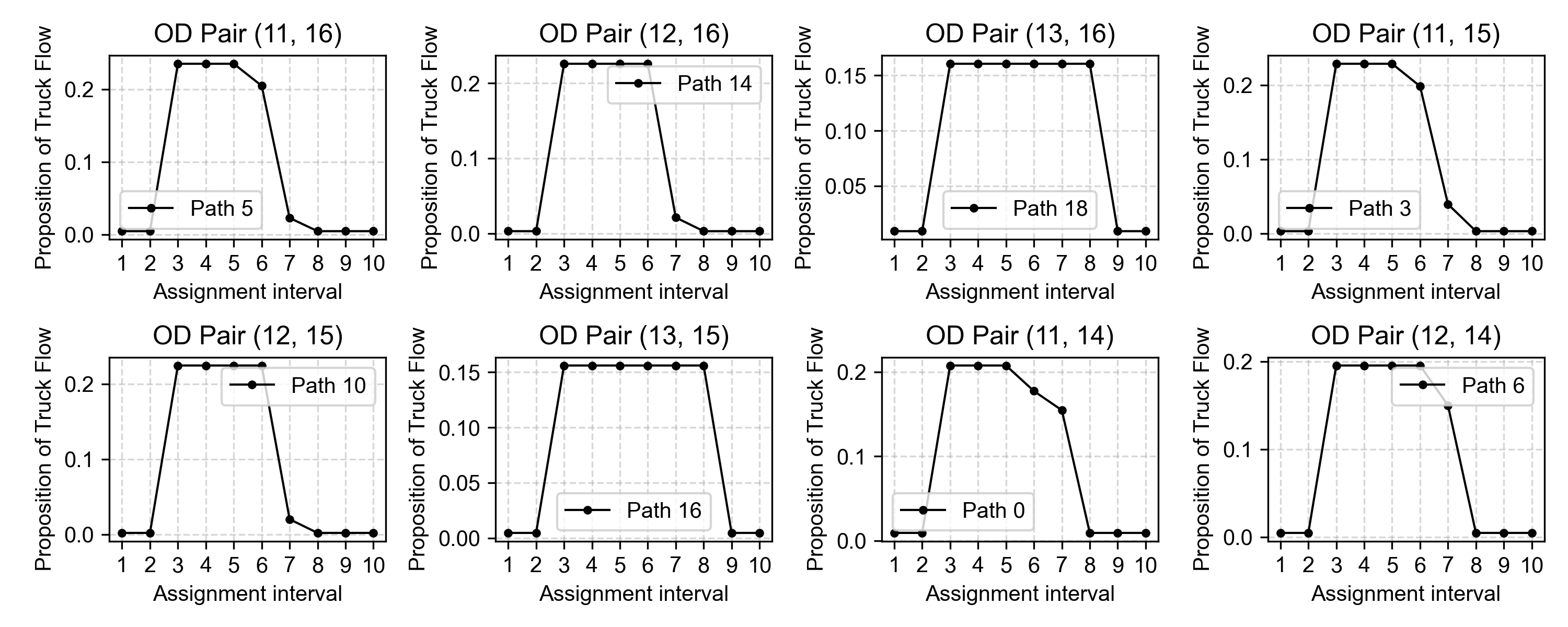}
    \caption{Truck path flow diversion (without inter-class PMCs)}
    \label{fig:od_diversion_dso_wo_truck}
\end{figure}

Figures~\ref{fig:od_diversion_dso_wo_car} and \ref{fig:od_diversion_dso_wo_truck} present the joint path and departure time choice results at DSO solved without considering inter-class PMCs. Unlike the inter-class PMC case, the results demonstrate {a very high degree of alignment in temporal path choice patterns across the two vehicle classes.} For nearly all OD pairs, cars and trucks follow similar path and departure time distributions, suggesting that using only intra-class PMCs can lead to {shared preferences in departure times and path choices that minimize each class's own travel costs, even if this exacerbates cross-class congestion.}

Another notable difference between two scenarios {lies} in the distribution of flows across assignment intervals. Without inter-class PMCs, several OD pairs exhibit a more even allocation across multiple assignment intervals, where cars and trucks select two or more intervals with similar proportions (e.g., OD pairs (13,16), (12,15), and (12,16)). In contrast, with inter-class PMCs, the flow distribution becomes significantly concentrated in one or two dominant intervals for most OD pairs. The introduction of inter-class terms steepens {PMCs} by accounting for cross-class externalities, effectively penalizing simultaneous usage by different classes during high-congestion periods. As a result, one or two intervals emerge as distinctly optimal, while other intervals become less attractive, leading to a sharper temporal concentration of flows.

\subsection{A large network}
\begin{figure}
    \centering
    \includegraphics[width=0.7\linewidth]{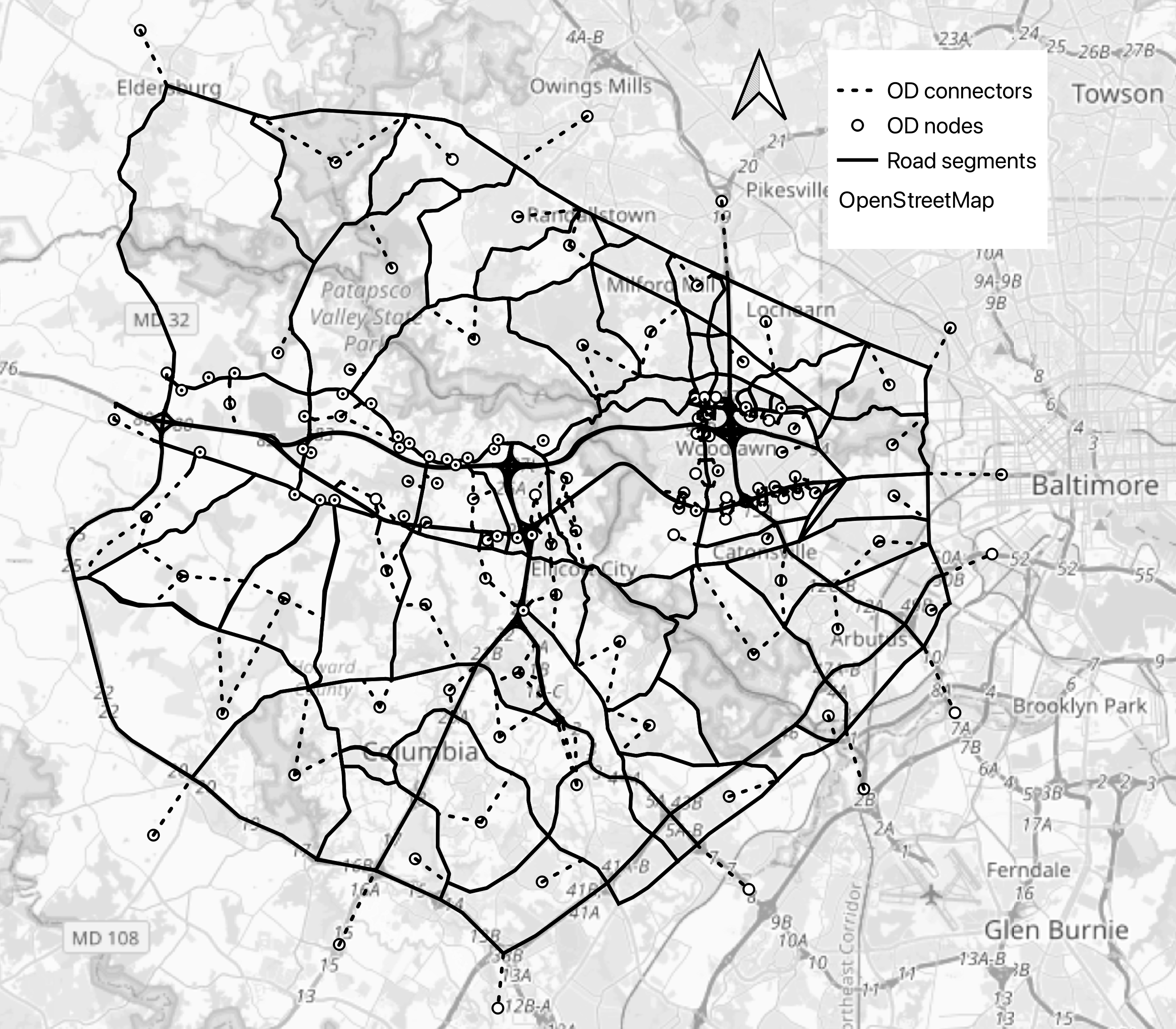}
    \caption{{The large-scale real-world network}}
    \label{fig:tsmo}
\end{figure}

To examine the scalability of the proposed method, we further solved DSO on a large-scale network with calibrated OD demand from real-world data. {The network is located near Baltimore, MD, provided by the Maryland Department of Transportation (MDOT) State Highway Administration (SHA).} The network contains 1,510 links, 776 nodes, 124 origins/destinations, and 15,376 OD pairs, as shown in Figure~\ref{fig:tsmo}. Detailed settings of the large network can be found in \citep{ke2023design, ke2025real}. The whole DNL models 7 hours, from 5 AM to 12 PM on a typical workday. Similarly, we consider both route and departure time choices and use the same travel cost parameters as the small network. We set the punctual time window to be 6:30-10:30 AM.

\begin{figure}[H]
    \centering
    \includegraphics[width=0.7\linewidth]{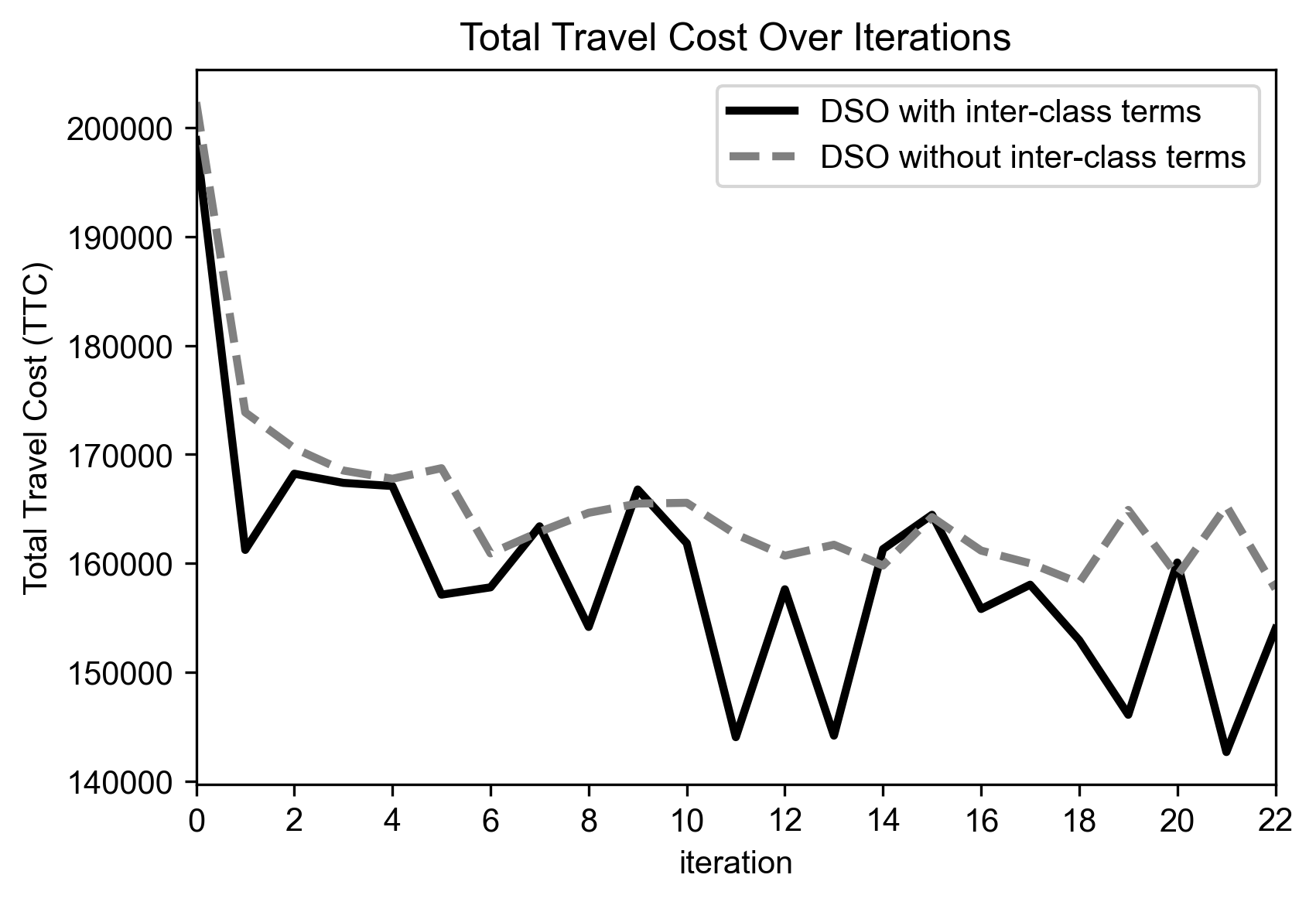}
    \caption{{Convergence of two DSO runs (with/without inter-class terms) on large network}}
    \label{fig:tsmo_converge}
\end{figure}

Figure~\ref{fig:tsmo_converge} shows the convergence curves of two DSO runs. We compare results using lower-bound PMC-based MSA with and without considering inter-class terms separately. The two DSO runs are considered to converge after 22nd iteration as the gap function values cannot be further reduced. The total gap values for DSO with and without inter-class terms achieve 0.674 and 0.894, respectively.

We observed that based on the same simple MSA algorithm, DSO with the consideration of inter-class terms yields a smaller total travel cost for the system. Its convergence curve continues to decrease after the 10th iteration, while DSO without inter-class terms maintains the same level of total cost and shows early convergence. We further decompose the total cost into different terms for cars and trucks in Table~\ref{tab:tsmo_res} and find that travel time costs of both classes are further reduced by 9\% and 5\%, respectively, when incorporating inter-class PMCs, indicating the potential to optimize mixed traffic flow patterns at the SO state.

\begin{table}[H]
\small
\centering
\begin{tabular}{lcccccc}
\hline
\textbf{Model Type} & \multicolumn{2}{c}{\textbf{Total Travel Time Cost (TTTC)}} & \multicolumn{2}{c}{\textbf{Total Travel Cost (TTC)}} & \multicolumn{2}{c}{\textbf{Gap}} \\
\cline{2-7}
& car        & truck      & car         & truck       & car         & truck       \\
\hline
with inter-class PMC     &  80820.3 & 7437.8 & 130936.8 & 11721.1 & 0.65 & 0.97\\
without inter-class PMC  & 89112.9 & 7812.1 & 145875.8 & 11770.9 & 0.92 & 0.37 \\
\hline
\end{tabular}
\caption{Decomposed travel cost by vehicle type in large network DSO (unit veh$\cdot$hour)}
\label{tab:tsmo_res}
\end{table}

{Compared to the small network, the convergence curve in the large network is less smooth and the final gap function value is relatively larger. This is mainly due to the higher problem dimension and the complex spatio-temporal interactions in a large network with bi-class vehicle setting. When inter-class externalities are considered, the algorithm must simultaneously balance the high-dimensional path and departure time choice of both vehicle classes, and oscillations can occur under vanilla MSA algorithm which simply depends on PMC values. We emphasize that the focus of this study is the proposed method of approximating heterogeneous PMCs and its effectiveness in providing meaningful descent directions for multi-class SO-DTA, rather than developing the most advanced large-scale SO-DTA solver. Using the same simple MSA algorithm, incorporating inter-class PMCs yields a lower total system travel cost and a smaller gap value than the baseline without inter-class terms. Developing stabilized or accelerated solution algorithms with improved step size rules and solvers tailored for large-scale networks to further reduce the final gap are important and promising directions for future work.}

\section{{Conclusion and Future Research}}
\label{sec:conclusion}
{This study develops a novel but simple analytical approach for approximately evaluating path marginal cost (PMC) in dynamic traffic networks with heterogeneous traffic flow. Accurately computing PMC under mixed traffic conditions is challenging because different vehicle classes interact and impose externalities on each other, and related computational issues arise when mixed traffic gets close to capacity. To address these challenges, this study decomposes PMC into intra-class and inter-class terms, and the inter-class term is quantified using a conversion factor derived from heterogeneous link dynamics, which provides a tractable way to represent how one class contributes to congestion experienced by the other class. In addition, we explicitly consider the non-differentiability that can occur near capacity in heterogeneous flow and incorporate a lower-bound (sub-gradient) PMC representation. These PMC approximations are applied directly in solving variational inequality formulations for system-optimal dynamic traffic assignment with standard iterative algorithms such as the method of successive averages.}

{We evaluate the proposed method on both a small corridor network with synthetic demand and a large-scale real-world network with calibrated demand from real-world data. Results across both settings reveal that incorporating inter-class PMC terms consistently improves the quality of the system optimum solution compared to the formulations that only include intra-class effects. In particular, the resulting system-optimal dynamic traffic assignment produces smoother and more efficient spatiotemporal flow distributions and better mitigates cross-class congestion externalities, with larger reductions in total travel cost and more competitive flow patterns at the system optimum. Additionally, it is shown that using lower-bound sub-gradient, which acknowledges the non-differentiable issue of heterogeneous flow, can further reduce total system cost, indicating that using a PMC approximation consistent with heterogeneous traffic dynamics is important for approaching the true system optimum state.}

{Solving multi-class path-based SO-DTA is one direct application enabled by heterogeneous PMCs developed in this study. 
The PMC formulation can also support a broader range of models for heterogeneous traffic networks, which are promising future research directions.
First, heterogeneous travel time estimation, multi-class traffic simulation calibration, and dynamic OD demand estimation can all benefit from heterogeneous PMC information, because these problems fundamentally rely on how changes in class-specific path flows propagate across links and paths through network traffic dynamics. In OD estimation, calibration and other inverse problems, gradient-based or sensitivity-informed updates are often required to efficiently match estimated states with real-world observations \citep{ma2020estimating, liu2023end, liu2025scalable, liu2025constructing, liu2025end}. From this perspective, heterogeneous PMCs provide a tractable approximation of these sensitivities under heterogeneous traffic condition, which helps improve both computational scalability and behavioral consistency when multiple vehicle classes coexist in the network model \citep{liu2024modeling, liu2024enhancing, liu2025curb}. When different classes interact asymmetrically (e.g., heavy vehicles impose disproportionate delays comparing small vehicles), using a homogeneous PMC approximation can bias the calibrated demand or parameters, but heterogeneous PMCs can explicitly consider these inter-class externalities and improve the estimation accuracy.}

{Second, heterogeneous PMCs naturally support deriving control strategies in heterogeneous network models, including pricing, signal control, and other interventions, where the objective function typically depends on travel time (or generalized travel cost) and the network dynamics serve as constraints. In such bi-level optimization problem formulations, we need to derive gradients of system performance with respect to decision variables that affect path flows. Some of the gradients can be expressed by marginal impacts of path flows on travel times, which is what PMCs approximate \citep{liu2023optimal}. Importantly, the proposed PMC decomposition into intra- and inter-class terms allows control variables to be class-aware, enabling control policy to distinguish whether a marginal change comes primarily from reducing intra-class congestion or from alleviating inter-class competition. This can be especially critical in heterogeneous problem settings \citep{wu2024participatory} where equity, prioritization, or operation are class-dependent.}

{Third, heterogeneous PMCs provide a useful metric for network resilience analysis, where the goal is to quantify how system performance changes when the network is perturbed and when mixed traffic composition shifts across classes. Conceptually, the PMC measures the marginal deterioration or improvement in system performance when adding or removing some path flows, making it as an explainable sensitivity metric to identify network vulnerable spots, critical paths, and the specific vehicle classes that contribute most to spatio-temporal congestion patterns. This class-specific PMC information is valuable for resilience evaluation under mixed traffic particularly when disruptions may amplify cross-class externalities (e.g., one class triggering congestion or spillover that disproportionately harms another), and heterogeneous PMCs can help model these inter-class impact mechanisms and support targeted proactive mitigation strategies.}

{Additionally, further research can focus on strengthening the heterogeneous PMC formulation and the solution algorithm for multi-class SO-DTA. First, the bi-class formulation can be generalized to multiple vehicle classes while preserving computational tractability for large-scale networks. Second, the lower-/upper-bound structure of the proposed PMCs motivates the development of more advanced non-smooth, subgradient-based algorithms beyond standard MSA to improve convergence speed and robustness in the presence of non-differentiability near capacity.}

\section*{Acknowledgments}
Part of this research was funded by National Science Foundation grant CMMI-2528901.
ChatGPT-5.2 was used to assist in improving the clarity, conciseness, and overall quality of the writing in this manuscript.

\bibliographystyle{apalike} 
\bibliography{reference}

\end{document}